\documentclass[12pt]{article}
\usepackage[width=180mm,top=20mm,bottom=20mm,bindingoffset=6mm]{geometry}
\usepackage[american]{babel}
\usepackage{amsmath,amssymb,amsfonts}
\usepackage{amsthm}
\usepackage{mathrsfs}
\usepackage{algorithm}
\usepackage{algorithmicx}
\usepackage{algpseudocode}
\usepackage{listings}
\usepackage{longtable}
\usepackage{setspace}
\usepackage[colorlinks=true, linkcolor=blue, urlcolor=blue, citecolor=blue]{hyperref}
\usepackage{orcidlink}
\usepackage[natbibapa]{apacite}
\usepackage[font=footnotesize,labelfont=bf]{caption}
\usepackage[font=footnotesize,labelfont=bf]{subcaption}

\newcommand{\parencite}[1]{\citep{#1}}

\begin{document}

	\pagenumbering{arabic}
	
	\setstretch{2.5}
	\begin{center}
		{\huge A new Timestep Criterion for the Simulation of Immiscible Two-Phase Flow with IMPES Solvers }
	\end{center}
	
	\vspace{0.5cm}	
	\onehalfspacing
	
	\begin{center}
		{ \large Dominik Burr$^{1,2}$ \orcidlink{0000-0003-1351-4994}, Dr. Stefan Rief$^{1}$, Dr. Konrad Steiner$^{1}$ }	
	\end{center}
	
	\begin{center}
		{ \large $^1$ Flow and Material Simulation, Fraunhofer ITWM, Kaiserslautern, Germany \\
			 	 $^2$ Department of Mathematics, RPTU Kaiserslautern-Landau, Kaiserslautern, Germany }
	\end{center}
	
	\begin{center}
		{ \large Corresponding Author: \href{mailto:dominik.burr@itwm.fraunhofer.de}{dominik.burr@itwm.fraunhofer.de} }	
	\end{center}
	
	\vspace{0.5cm}	

	{ \large { \centering \textbf{Abstract} \\ } }
	
	We present an IMPES solver and a novel timestep criterion for the simulation of immiscible two-phase flow involving compressible fluid phases. The novel timestep criterion uses the Courant-Friedrichs-Lewy (CFL) condition and employs numerically computed velocity derivatives to adapt the timestep size, regardless of the dominant flow characteristics. \\
	The solver combined with this timestep criterion demonstrates both efficiency and robustness across a range of flow scenarios, including pressure drop dominated and capillary dominated flows with compressible and incompressible fluid phases, without the need to adjust any numerical parameters. Furthermore, it successfully reaches the expected stationary states in a case involving discontinuous porous media parameters such as porosity, permeabilities, and capillary pressure function.  \\
	Comparison with the established Coats timestep criterion reveals that our approach requires fewer time iterations while maintaining comparable accuracy on the Buckley-Leverett problem and a gravity-capillary equalization example with a known stationary state. Additionally, in an example with air compression, the new timestep criterion leads to a significantly improved non-wetting phase mass conservation compared to the Coats criterion. \\

	
	\newpage

	\section{Introduction}
		
		Simulating two-phase flow in porous media plays a crucial role across a variety of fields. In geological engineering, it supports applications such as extraction of gas and oil \parencite{mouketou2019modelling, jiang2018coupled}, investigations of groundwater contamination \parencite{myers2012potential, bastian1999numerical}, geological CO2 storage \parencite{march2018assessment, ren2017two}, and analysis of salt precipitation \parencite{schollenberger2025two}. Outside of geology, the simulation of two-phase flow is for instance used to simulate manufacturing processes including electrolyte injection in lithium-ion batteries \parencite{hagemeister2022numerical, gunter2022influence} and the fabrication of composite materials through techniques such as liquid composite molding \parencite{zhao2019three, michaud2016review}. Additionally, two-phase flow modeling has been utilized to study transport phenomena in fuel cells by \cite{sun2009fast} and \cite{djilali2008transport}. \\

		We describe immiscible two-phase flow using the two-phase Darcy equations for Newtonian compressible fluids. This is a non-linear partial differential equation (PDE) system and it is introduced in Section \ref{section_darcyEq}. \\
		Two primary approaches exist for simulating this equation system. The first is the fully implicit method, which solves the entire coupled PDE system implicitly \parencite{cao2002development, jiang2017efficient} and is unconditionally stable. The second is the sequential approach, encompassing IMplicit Pressure Explicit Saturation (IMPES) methods \parencite{horgue2015open, chen2019fully, redondo2018efficiency} and iterative schemes \parencite{pau2009parallel, jenny2006adaptive, el2019iterative}. Here, the PDE system is decoupled and solved sequentially. This reduces the size of the linear system solved at each timestep, lowering memory usage and runtime per time iteration. In IMPES methods, the saturation equation is solved explicitly. Consequently, only for the simulation of the pressure equation a linear system of equations needs to be solved. However, sequential methods are only conditionally stable. That means the timestep size needs to be restricted to maintain the stability of the numerical solver. As a consequence, the computational savings per time iteration may be lost by the need for smaller timestep sizes. \\
		
		While fully implicit methods generally handle compressible fluid phases without difficulty, sequential methods face greater challenges because fluid phase masses depend explicitly on phase pressure and saturation. Since the pressure and saturation equations are solved sequentially, errors from this decoupling can lead to violations of the mass conservation of the fluid phases. \\
		\cite{chi2025effect} developed an IMPES method for two-phase flow with a compressible non-wetting phase, excluding capillary forces. Sequential methods that discretize the saturation equation implicitly or semi-implicitly for compressible two-phase flow are for instance presented by \cite{pau2012adaptive}, \cite{lu2009iterative}, and \cite{lee2008multiscale}. \\
		Because sequential methods are only conditionally stable, selecting appropriate timestep sizes is critical. Several timestep criteria specifically for sequential immiscible two-phase flow simulations exist. In Section \ref{sec_charWaveVel}, we mention two such criteria from the literature. The first, known as the Coats criterion, is derived via Neumann stability analysis \parencite{coats2003impes}. The second, termed by us as the characteristic wave velocity criterion, is used by \cite{lamine2015multidimensional} and is based on the Courant-Friedrichs-Lewy (CFL) condition \parencite{moura2012courant}. \\

		Section \ref{section_numerics} presents an IMPES solver with a finite volume discretization, along with a novel timestep criterion. This novel timestep criterion generalizes the characteristic wave velocity criterion, which, to the knowledge of the authors, is new to the literature. The timestep criterion uses a numerical approximations of the derivative of the wetting phase velocity to limit the timestep based on the CFL condition. A detailed description is provided in Section \ref{sec_charWaveVel}. By incorporating capillary forces and numerical approximations of saturation derivatives of velocities, this generalized criterion effectively restricts the timestep regardless of the dominant flow dynamics. Throughout this work, it is referred to as the generalized characteristic wave velocity criterion. \\

		The solver combined with the timestep criterion is evaluated and validated through several examples in Section \ref{sec:numericalExperiments}, including capillary dominated and pressure drop dominated flows. Additionally, the capability of the solver to accurately simulate compressible fluid phases and to handle flows in geometries with discontinuous material parameters is assessed. Simulations involving discontinuous material parameters are of interest in the literature as there are various applications of two-phase flow with discontinuous material parameters \parencite{ma2021discontinuous, helmig1998comparison, bastian2014fully, brenner2013finite}. In the considered flow examples, the accuracy and efficiency of the new timestep criterion are compared against the Coats and the characteristic wave velocity criteria. These comparisons indicate that the new timestep criterion is better suited to restrict the timestep of capillary dominated flows than the characteristic wave velocity criterion. Moreover, compared to the Coats criterion, it reduces the number of time iterations required while maintaining similar accuracy, and in the example from Section \ref{sec:darcySol_compressGas}, it achieves a significantly better mass conservation of the air.

	\section{Materials and Methods}
	
		\subsection{Two-Phase Darcy Equations}  \label{section_darcyEq}

			The two-phase Darcy equations for Newtonian fluids read
			\begin{flalign} \label{two_phase_darcy_comp_2fluid_satEq}
				\frac{\partial \phi \rho_d S_d}{\partial t} &= - \nabla \cdot (\rho_d u_d) \; \; \text{for } d \in \{ w, n \} ,  \\ \label{two_phase_darcy_comp_2fluid_velEq}
				u_d &= - M_d K_0 \left( \nabla p_d - \rho_d g \right) \; \; \text{for } d \in \{ w, n \} , \\ \label{two_phase_darcy_comp_2fluid_satConst}
				S_w + S_n &= 1 ,  \\ \label{two_phase_darcy_comp_2fluid_capPress}
				p_n - p_w &= p_c(S_w) .
			\end{flalign}
			The subscripts "w" and "n" represent the wetting and non-wetting phase, respectively. Both fluid phases and the solid phase involved determine whether a fluid is the wetting or non-wetting phase. The unknowns of the PDE system are usually the phase velocities $u_d$, the phase pressures $p_d$ and the phase saturations $S_d$. The phase saturations $S_d \in [0, 1]$ indicate how much pore space is occupied by phase $d$. \\	
			The phase mobilities $M_d$ are defined as
			\begin{equation}
				M_d = \frac{k_{rd}}{\mu_d} \; \; \text{for } d \in \{ w, n \} ,
			\end{equation}
			and the total mobility $M$ as
			\begin{equation}
				M = M_w + M_n  .
			\end{equation}
			Moreover, $g$ is the gravity constant, $\phi \in [0, 1]$ is the porosity, $K_0$ is the absolute permeability tensor, $\rho_d$ is the density of phase $d$ and $\mu_d$ is the viscosity of phase $d$. \\
			The capillary pressure function $p_c$ and the relative permeabilities $k_{rd}$ are material parameters that are dependent on the phase saturations $S_d$. The capillary pressure function relates the pressure of the wetting phase to the pressure of the non-wetting phase by Eq. \ref{two_phase_darcy_comp_2fluid_capPress}. The relative permeabilities are scalar functions with values between zero and one that model the reduced ability of a fluid phase to flow through a porous medium when its pores are partially occupied by another fluid. \\	
			
			In all examples of this work, we assume either constant phase densities or the ideal gas law \parencite{laugier2007derivation}. Because we assume the temperature to be constant during our simulations, the ideal gas law states that there is a linear dependence between the density and the pressure of a fluid phase. We denote 
			\begin{equation} \label{idealGasLaw}
				\rho_d = \frac{1}{R_d} (p_d - p_d^{ref}) \rho_d^{ref} + \rho_d^{ref} ,
			\end{equation}
			where $R_d$ is the gas constant of phase $d$ and $\rho_d^{ref}$ is the density of the phase at the reference phase pressure $p_d^{ref}$. To evaluate Eq. \ref{idealGasLaw} the value of $R_d$ and the density of the fluid phase at one pressure value are required. \\

			To reduce the number of unknowns, we transform the equation system \ref{two_phase_darcy_comp_2fluid_satEq} - \ref{two_phase_darcy_comp_2fluid_capPress} to
			\begin{flalign} \label{two_phase_darcy_comp_satEq}
				\frac{\partial \phi \rho_w S_w}{\partial t} &= - \nabla \cdot \left( f_w \rho_w \left( u + M_n u_D \right) \right), \\ \label{two_phase_darcy_comp_velEq}
				u &= K_0 (-M \nabla p_n + M_w \nabla p_c + (M_w \rho_w + M_n \rho_n) g), \\ \label{two_phase_darcy_comp_pressEq}
				0 &= \nabla \cdot u + \sum_{d=w,n} \frac{1}{\rho_d} \left( \nabla \rho_d \cdot u_d + \phi S_d \frac{\partial \rho_d}{\partial t} \right),
			\end{flalign}
			where
			\begin{flalign} \label{u_w}
				u_w &= f_w u +  f_w K_0 M_n \left( \nabla p_c + (\rho_w - \rho_n) g \right) = f_w u + \gamma u_D , \\ \label{u_n}
				u_n &= f_n u - f_w K_0 M_n \left( \nabla p_c + (\rho_w - \rho_n) g \right) = f_n u - \gamma u_D , \\ \label{u_D}
				u_D &= K_0 \left( \nabla p_c + (\rho_w - \rho_n) g \right), \\
				u &= u_w + u_n, \\
				p_c &= p_n - p_w .
			\end{flalign}
			In this formulation $u$ is the so-called total velocity. Further, it is
			\begin{flalign}
				f_d &= \frac{M_d}{M}, \\
				\gamma &= \frac{M_w M_n}{M} = M_n f_w  .
			\end{flalign}
			A similar reformulation of the two-phase Darcy equations was done by \cite{bastian1999numerical}.

		\subsection{Capillary Pressure and Relative Permeability Functions} \label{section_capPressAndRelPermeab}
		
		In this work, we use simplified Brooks-Corey relative permeabilities and as capillary pressure functions we either use Brooks-Corey or Van Genuchten models. \\
		To introduce these, we first define effective phase saturations by
		\begin{flalign*}
			\bar{S}_d = \frac{S_d - S_{d,r}}{1 - S_{w,r} - S_{n,r}} \; \text{ for } d = w, n .
		\end{flalign*}
		In the above equation $S_{w,r}$ is the residual wetting phase saturation and $S_{n,r}$ is the residual non-wetting phase saturation. \\
		The Brooks-Corey capillary pressure function \parencite{brooks1965hydraulic} is given by
		\begin{flalign*}
			p_c^{BC} (S_w) = p_e^{BC} \bar{S}_w^{- m^{BC}},
		\end{flalign*}
		where $p_e^{BC}$ and $m^{BC}$ are parameters that are depended on the porous medium. \\
		The Van Genuchten capillary pressure function \parencite{van1980closed} is given by
		\begin{flalign*}
			p_c^{VG} (S_w) = p_e^{VG} \left( \left( \bar{S}_w \right)^{- \frac{1}{m^{VG}}} - 1 \right)^{1 - m^{VG}} .
		\end{flalign*}
		Here $p_e^{VG}$ and $m^{VG}$ need to be adapted to the respective porous medium. \\	
			
		As relative permeability models, we use the simplified Brooks-Corey functions \parencite{corey1994mechanics} that are given by
		\begin{flalign*}
			k_{rw}^{BC} (S_w) &= \left( \bar{S}_w \right)^{m^{w,BC}}, \\
			k_{rn}^{BC} (S_w) &= \left( 1 - \bar{S}_w \right)^{m^{nw,BC}} ,
		\end{flalign*}
		where $m^{w,BC}$ and $m^{nw,BC}$ are material parameters. In Section \ref{sec_charWaveVel}, we formulate a maximization problem involving the derivatives of the relative permeabilities with respect to the saturation $S_w$. When $m^{w,BC} \geq 1$ and $m^{n,BC} \geq 1$, this poses no issue, because the derivatives of the Brooks-Corey relative permeabilities are finite. However, if either $m^{w,BC} < 1$ or $m^{n,BC} < 1$, the derivatives become unbounded. In this case, it is necessary to restrict the derivatives during the maximization.

		\subsection{IMPES Solver}  \label{section_numerics}
		
			The solution of the PDE system (\ref{two_phase_darcy_comp_satEq} - \ref{two_phase_darcy_comp_pressEq}) is approximated using an IMPES method on a voxel grid. Moreover, a timestep criterion is provided to ensure the stability of the solver. \\
			Although this work only considers the ideal gas law or incompressible fluid phases in the examples, the solver presented in this section can, in principle, accommodate other relations between phase pressures and phase densities. The necessary inputs for the solver are the function describing this relation and its derivative with respect to the respective phase pressure. We remark that such cases are not included in our validation examples.
			
			\begin{figure}[ht]
				\centering
				\includegraphics[width=0.6\textwidth]{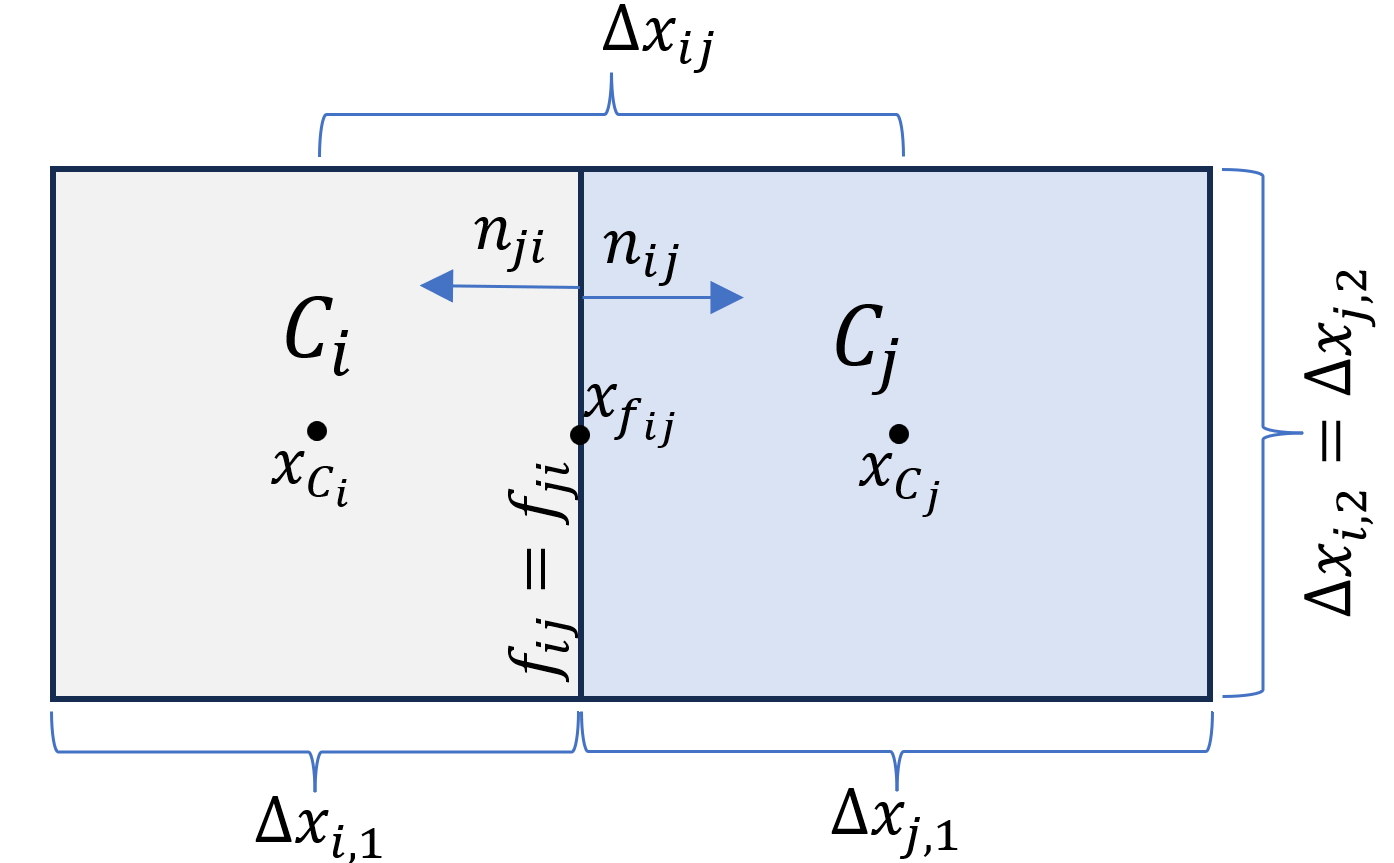} 
				\caption{This is a cutout of a two-dimensional grid with the relevant grid parameters as defined in Section \ref{section_numerics}. }
				\label{cellsAndFace}
			\end{figure}
			
			An individual cell of the voxel grid is denoted by $C_i$, where $i \in \mathbb{N}$. All cells are cuboids or rectangles, dependent if the grid is two- or three-dimensional, that are aligned with the canonical unit vectors $e_d$ for $d=1,...,n_D$. We refer to cells similarly as voxels. We denote the dimension of the space by $n_D \in \{ 2, 3 \}$. We explain the notations of the grid only for the three-dimensional case. The notations in the two-dimensional case are accordingly. \\
			
			The edge length of a cell $C_i$ is denoted by $\Delta x_{i,d}$ and the volume of the cell by $|C_i|$. Because the cells are cuboids, it is 
			\begin{equation*}
				|C_i| = \prod_{d=1}^{n_D} \Delta x_{i,d} .
			\end{equation*}
			With $f_{ij}$, we denote a shared face of the voxels $C_i$ and $C_j$. We denote the face normal of $f_{ij}$ pointing into $C_j$ as $n_{ij}$ and the face normal pointing into $C_i$ as $n_{ji}$. As the grid is aligned to the canonical unit vectors we can define
			\begin{flalign*}
				d_{ij} \in \{ 1,...,n_D \} \; \text{ so that } \; n_{ij} = - n_{ji} = \pm e_{d_{ij}} .
			\end{flalign*}
			Further, we define the function 
			\begin{equation*}
				\text{sign} (n_{ij}) := \left\{\begin{array}{ll} 
					+1, \; \; \; &\text{if } n_{ij} = e_{d_{ij}} \;  \\
					-1, &\text{if } n_{ij} = -e_{d_{ij}} \;   \end{array} \right. .
			\end{equation*}
			
			With $\mathbb{V}$, we denote the set of all cells of the grid. By $F_{C_i}$ , we denote the set containing all faces of the cell $C_i$, and by $F_{C_i}^{d}$, we denote the set containing all faces of the cell $C_i$ that are orthogonal to $e_d$. There are always exactly two faces in the sets $F_{C_i}^{d}$. \\
			We denote the center of a cell $C_i$ by $x_{C_i}$ and the center of a face $f_{ij}$ by $x_{f_{ij}}$. The distance between $x_{C_i}$ and $x_{C_j}$ is denoted by $\Delta x_{ij}$ and the area of a face $f_{ij}$ is denoted by $|f_{ij}|$. \\
			A two-dimensional example of all the defined grid variables is given in Figure \ref{cellsAndFace}. \\
			
			In the solver it is necessary to evaluate variables on face centers which only have values on cell centers. Therefore, averages of the adjacent cell-centered values are used. To simplify the notation in the following sections, we define for an arbitrary variable $\psi$ the harmonic and arithmetic average on a face $f_{ij}$ by
			\begin{flalign} \label{eq:darcySol_harmMeanFace}
				\overline{\psi}^{H,f_{ij}} &:= \frac{2 \Delta x_{ij} \psi_i \psi_j }{\Delta x_{i,d_{ij}} \psi_j + \Delta x_{j,d_{ij}} \psi_i } , \\ \label{eq:darcySol_arithMeanFace}
				\overline{\psi}^{A,f_{ij}} &:= \frac{1}{2 \Delta x_{ij}} \left( \Delta x_{i,d_{ij}} \psi_j + \Delta x_{j,d_{ij}} \psi_i \right) .
			\end{flalign}
			
			In the solver, it is necessary to approximate the gradients of variables, such as the capillary pressure or the non-wetting phase pressure. We always approximate these at face centers in the form 
			\begin{flalign} \label{eq:darcySol_faceGrad}
				\nabla q_{f_{ij}} \cdot n_{ij} := \frac{ \text{sign} (n_{ij}) }{\Delta x_{ij} } \left( q_{j} - q_{i}  \right) ,
			\end{flalign}
			for an arbitrary variable $q$ and a face $f_{ij}$ between the cells $C_i$ and $C_j$.

		\subsubsection{Pseudo Algorithm} \label{section_pseudoalg}
			Below the finite volume based method to solve the two-phase Darcy equations is given as pseudo code. It is assumed that the first $k-1$ time iterations were completed and that the variables $S_{w,i}^{k-1}, S_{w,i}^{k-2}, p_{n,i}^{k-1},$ $p_{n,i}^{k-2}, u_{f_{ij}}^{k-1}, u_{D,f_{ij}}^{k-1}, u_{i}^{k-1}, u_{i}^{k-2}, u_{D,i}^{k-1}$ and $ u_{D,i}^{k-2}$ are accessible to the algorithm. The index $i$ refers to the cell $C_i$ in the grid, the index $k$ denotes the timestep number, and $f_{ij}$ denotes the face between cells $C_i$ and $C_j$ in the grid. \\
			The parameter $m_I \geq 1$ refers to the number of IMPES iterations, i.e., the number of sequential solves of the PDE system that is carried out in every time iteration. The number $m_I$ is fixed in the algorithm, meaning that in each timestep exactly $m_I$ IMPES iterations are performed, indexed by $l=1,...,m_I$.
			
			\begin{enumerate} 
				\item Select the timestep $\Delta t_k$. \label{alg_timestep}
				\item For $l=1,..., m_I$ repeat: \\
				Define $k_{0} := k-1$.
				\begin{enumerate}
					\item Calculate the saturation $S_{w,i}^{k_l}$ on all cells $C_i$.  \label{alg_saturation}
					\item Calculate the pressure $p_{n,i}^{k_l}$ on all cells $C_i$. \label{alg_pressure}
					\item Calculate the velocities $u_{f_{ij}}^{k_l}$ and $u_{D,f_{ij}}^{k_l}$ on all faces $f_{ij}$. \label{alg_faceVelocities}
				\end{enumerate}
				After the last IMPES iteration $l = m_I$: \\
				Define $S_{w,i}^{k} := S_{w,i}^{k_{m_I}}$, $p_{n,i}^{k} := p_{n,i}^{k_{m_I}}$, $u_{f_{ij}}^{k} := u_{f_{ij}}^{k_{m_I}}$ and $u_{D,f_{ij}}^{k} := u_{D,f_{ij}}^{k_{m_I}}$ .
				\item Calculate the cell velocities $u_{i}^{k}$ and $u_{D,i}^{k}$ on all cells $C_{i}$ by linear interpolation of the respective face velocities.  \label{alg_cellVelocities}
				
				\item Continue with the next time iteration $k+1$.
			\end{enumerate}
			
			Within the inner loop for $l=1, \ldots, m_I$, it is required to store $S_w$ and $p_n$ from both the previous timestep $k-1$ and the last IMPES iteration $k_{l-1}$. However, for the velocities $u_{f_{ij}}$ and $u_{D,f_{ij}}$, only the values from the latest IMPES iteration $k_{l-1}$ need to be stored. \\			
			
			In the proceeding sections the individual steps of the algorithm are explained in detail. The selection of the timestep, Step \ref{alg_timestep}, is explained after the other steps of the algorithm.

		\subsubsection{Calculation of the Saturation} \label{sec_satEq}
		In this section, Step \ref{alg_saturation} of the method is explained in more detail. \\	
		We rewrite the left-hand side of Eq. \ref{two_phase_darcy_comp_satEq} as
		\begin{equation*}
			\frac{\partial \phi \rho_w S_w}{\partial t} = \phi \left( \rho_w \frac{\partial S_w}{\partial t} + S_w \frac{\partial \rho_w }{\partial t} \right) = \phi \left( \rho_w \frac{\partial S_w}{\partial t} + S_w \frac{\partial \rho_w }{\partial p_w} \left( \frac{\partial p_n }{\partial t} - \frac{\partial p_c }{\partial t} \right) \right) .
		\end{equation*}
		We use a Euler scheme to discretize the time derivative of $S_w$. The explicit occurrences of $S_w$ are implicitly discretized using $S_w^{k_l}$, i.e., the saturation that is determined in this step. On the right-hand side of Eq. \ref{two_phase_darcy_comp_satEq}, we use the Gauss divergence theorem and an upwind scheme. In the material functions the saturation values of the last IMPES iteration are used. This leads to the discretized equation
		\begin{equation} \label{discreteSatEq}
			S^{k_l}_{w,i} = \frac{ S^{k-1}_{w,i} - \frac{\Delta t_k }{\phi_i |C_i| \rho_{w,i}^{k_{l-1}}} \sum_{f_{ij} \in F_{C_i}} |f_{ij}| \rho_{w,f_{ij}}^{k_{l-1}}  F^{up, k_{l-1}}_{f_{ij}} }{ 1 + \frac{1}{\rho_{w,i}^{k_{l-1}}} \frac{\partial \rho^{k_{l-1}}_{w,i}}{\partial p_w} \left( p_{n,i}^{k_{l-1}} - p_{c,i}^{k_{l-1}} - p_{n,i}^{\tilde{k}_l} + p_{c,i}^{\tilde{k}_l} \right) } ,
		\end{equation}
		of the saturation in cell $C_i$. In this formula the index
		\begin{flalign} \label{eq:Tildek_l}
			\tilde{k}_l &= \left\{\begin{array}{ll} k-2, \; \; \; \; \; &\text{if} \; l = 1 \\
				k-1, &\text{else} \end{array} \right. ,
		\end{flalign}
		must be defined differently for the first IMPES iteration in a time iteration. Otherwise, the denominator in Eq. \ref{discreteSatEq} would be $0$ in the first IMPES iteration. \\	
		The numerical flux function is defined as 	
		\begin{flalign} \label{numericalFluxFct}
			F^{up, k_{l-1}}_{f_{ij}} &= f^{up, k_{l-1}}_{w,f_{ij}} \left( u^{k_{l-1}}_{f_{ij}} + \tilde{u}^{k_{l-1}}_{D,f_{ij}} \right)  \cdot n_{ij} ,
		\end{flalign}
		where the coefficient function $f_w$ is calculated by an upwind scheme
		\begin{flalign} \label{upwinding_fw}
			f^{up, k_{l-1}}_{w,f_{ij}} &:=  \left\{\begin{array}{ll} 
				f_{w} \left( S_{w,i}^{k_{l-1}} \right), &\text{if} \; \; \left( u^{k_{l-1}}_{f_{ij}} + \tilde{u}^{k_{l-1}}_{D,f_{ij}} \right)  \cdot n_{ij} \geq 0 \\
				f_{w} \left( S_{w,j}^{k_{l-1}} \right), &\text{else} 
			\end{array} \right. .
		\end{flalign}
		Moreover, it is
		\begin{flalign} \label{darcySol_numFluxFct}
			\tilde{u}^{k_{l-1}}_{D,f_{ij}} := \overline{M_n^{k_{l-1}} K_0}^{H,f_{ij}} \nabla p_{c,f_{ij}}^{k_{l-1}} + \overline{M_n^{k_{l-1}} K_0}^{H,f_{ij}} \left( \overline{ \rho^{k_{l-1}}_w }^{A,f_{ij}} - \overline{ \rho^{k_{l-1}}_n }^{A,f_{ij}} \right) g .
		\end{flalign}
		How the total velocity $u^{k_{l-1}}_{f_{ij}}$ is calculated, is explained in Section \ref{sec_faceVel}. \\
				
		In Eq. \ref{darcySol_numFluxFct}, we select the harmonic mean of $M_n K_0$ to ensure consistency with the discrete pressure equation. The discrete pressure equation is presented in Section \ref{sec_pressEq}. Maintaining consistency between the discrete saturation and pressure equations is essential because, in the formulation of the two-phase Darcy equations used by the solver, mass conservation of the non-wetting phase is only implicitly guaranteed by combining the saturation and pressure equation. Furthermore, since compressible fluid phases are considered, updating the phase pressure also influences the phase masses. If now for instance the total velocity is discretized differently in the pressure and saturation equations, this could introduce an error in the non-wetting phase mass. \\
		Therefore, the harmonic mean of the phase mobilities times the absolute permeability is consistently applied throughout all steps of the algorithm. For the same reason the upwinding of $f_w$ for calculating $u_w$ in Eq. \ref{darcySol_numFluxFct} is also consistently used for discretizing the phase velocities $u_d$ in the solver. \\
		If the timesteps are small enough, the upwinding and the definition of the discretized velocities ensure that the saturation cannot exceed the bounds given by the irreducible phase saturations.

		\subsubsection{Calculation of the Pressure}
		 \label{sec_pressEq}

		In this section, Step \ref{alg_pressure} of the method is explained in more detail. \\
		By using 
		\begin{flalign*}
			\nabla \cdot \left( \rho_d u_d \right) = \nabla \rho_d \cdot u_d + \rho_d \nabla \cdot u_d \; \Leftrightarrow \;  \nabla \rho_d \cdot u_d = \nabla \cdot \left( \rho_d u_d \right) - \rho_d \nabla \cdot u_d ,
		\end{flalign*}
		Eq. \ref{two_phase_darcy_comp_pressEq} can be rewritten as
		\begin{flalign} \label{tmpPressEq}
			0 &= \nabla \cdot u + \sum_{d=w,n} \frac{1}{\rho_d} \left( \nabla \cdot \left( \rho_d u_d \right) - \rho_d \nabla \cdot u_d + \phi S_d \frac{\partial \rho_d}{\partial t} \right) .
		\end{flalign}
		We employ this reformulation as this allows us to use only discretizations of gradients on faces. Before the reformulation, we would need to calculate gradients on cells for the discretizations of the phase velocities $u_d$ on cells. \\
		We do not want to discretize gradients on cells as this can quickly lead to checkerboard instabilities. This is as in the straightforward discretization of a cell-centered gradient, the discrete gradient is not dependent on the variable value on the same cell. If this approximation is used in a finite volume solver the cells are separated in two disjoint sets of cells, arranged in a checkerboard pattern, which only interact among themselves, and easily lead to instabilities. This behavior was analyzed for the approximation of the Navier-Stokes equations by \cite{rhie1983numerical} and for the approximation of the wave equation by \cite{dellacherie2009checkerboard}. \\

		To simplify the notation in the following, we define 
		\begin{flalign} \label{u_R}
			u_R &:= u + M K_0 \nabla p_n = K_0 \left( M_w \nabla p_c \left( M_w \rho_w + M_n \rho_n \right) g \right) \\ \notag
			\Leftrightarrow u &= - M K_0 \nabla p_n +  u_R .
		\end{flalign}
		By inserting this, together with Eq. \ref{u_w} and Eq. \ref{u_n}, into Eq. \ref{tmpPressEq} and rearranging terms, we get
		\begin{flalign*}
			\sum_{d=w,n} \frac{1}{\rho_d} \nabla \cdot & \left( \rho_d f_d M K_0 \nabla p_n \right) - \left( \frac{\phi S_w}{\rho_w} \frac{\partial \rho_w}{\partial p_w } + \frac{\phi S_n}{\rho_n} \frac{\partial \rho_n}{\partial p_n } \right) \frac{\partial p_n}{\partial t} = \sum_{d=w,n} \frac{1}{\rho_d} \nabla \cdot \left( \rho_d f_d u_R \right)  \\
			&+ \frac{1}{\rho_w} \nabla \cdot \left( \rho_w f_w M_n u_D \right) -  \frac{1}{\rho_n} \nabla \cdot \left( \rho_n f_w M_n u_D \right) - \frac{\phi S_w}{\rho_w} \frac{\partial \rho_w}{\partial p_w} \frac{\partial p_c}{\partial t} .
		\end{flalign*}
		We discretize this equation by integrating over a cell $C_i$ and using the Gauss divergence theorem. The discrete form of the equation that we use is 
		\begin{flalign} \notag
			& \sum_{f_{ij} \in F_{C_i}} |f_{ij}| A^{k_l}_{f_{ij}} \overline{M^{k_l} K_0 }^{H,f_{ij}} \nabla p^{k_l}_{n,f_{ij}} \cdot n_{ij} - \frac{|C_i| \phi_i }{\Delta t_k } \left( \frac{S^{k_l}_{w,i}}{\rho^{k_{l-1}}_{w,i}} \frac{\partial \rho^{k_{l-1}}_{w,i} }{\partial p_w} + \frac{(1 - S^{k_l}_{w,i})}{\rho^{k_{l-1}}_{n,i}} \frac{\partial \rho^{k_{l-1}}_{n,i}}{\partial p_n} \right) p^{k_l}_{n,i} \\ \label{discretePressEq}
			&= \sum_{f_{ij} \in F_{C_i}} |f_{ij}| A^{k_l}_{f_{ij}} \hat{u}^{k_l}_{R, f_{ij}} \cdot n_{ij} + \sum_{f_{ij} \in F_{C_i}} |f_{ij}| B^{k_l}_{f_{ij}} f^{up, k_l}_{w,f_{ij}} \hat{u}^{k_l}_{D, f_{ij}}   \cdot n_{ij} \\ \notag
			&- \frac{|C_i| \phi_i }{\Delta t_k } \left( \frac{S^{k_l}_{w,i}}{\rho^{k_{l-1}}_{w,i}} \frac{\partial \rho^{k_{l-1}}_{w,i} }{\partial p_w} \left( p^{k-1}_{n,i} + p_{c,i}^{k_l} - p_{c,i}^{k-1} \right) + \frac{(1 - S^{k_l}_{w,i})}{\rho^{k_{l-1}}_{n,i}} \frac{\partial \rho^{k_{l-1}}_{n,i}}{\partial p_n} p^{k-1}_{n,i} \right),
		\end{flalign}
		where
		\begin{flalign} \label{discretA}
			A^{k_l}_{f_{ij}} &:= \left( f^{up,k_{l}}_{w,f_{ij}} \frac{\rho_{w,f_{ij}}^{k_{l-1}}}{\rho_{w,i}^{k_{l-1}}}  + \left( 1 - f^{up,k_{l}}_{w,f_{ij}} \right) \frac{\rho_{n,f_{ij}}^{k_{l-1}}}{\rho_{n,i}^{k_{l-1}}}  \right) , \\ \label{discretB}
			B^{k_l}_{f_{ij}} &:= \frac{\rho_{w,f_{ij}}^{k_{l-1}}}{\rho_{w,i}^{k_{l-1}}} - \frac{\rho_{n,f_{ij}}^{k_{l-1}}}{\rho_{n,i}^{k_{l-1}}} , \\ \label{discretu_R}
			\hat{u}^{k_l}_{R, f_{ij}} &:= \overline{M_w^{k_{l}} K_0 }^{H,f_{ij}} \nabla p_{c,f_{ij}}^{k_l} + \left( \overline{K_0 M_w^{k_l}}^{H,f_{ij}} \overline{\rho^{k_{l-1}}_w}^{A,f_{ij}} + \overline{K_0 M_n^{k_l}}^{H,f_{ij}} \overline{\rho^{k_{l-1}}_n}^{A,f_{ij}} \right) g , \\ \label{discretu_D}
			\hat{u}^{k_l}_{D, f_{ij}} &:= \overline{M_n^{k_l} K_0}^{H,f_{ij}} \nabla p_{c,f_{ij}}^{k_l} + \overline{M_n^{k_l} K_0}^{H,f_{ij}} \left( \overline{\rho^{k_{l-1}}_w}^{A,f_{ij}} - \overline{\rho^{k_{l-1}}_n}^{A,f_{ij}}  \right) g .
		\end{flalign}
		We solve the resulting linear system of equations by an algebraic multigrid method \parencite{krechel1998operator}. \\

		To justify the usage of the harmonic mean for the discretization of the velocities, we consider the phase velocity $u_d$ as given by Eq. \ref{u_w} or Eq. \ref{u_n} respectively. We neglect the gravity term and we assume the phase velocity $u_d$ as well as the phase pressure $p_d$ to be continuous between two neighboring cells $C_i$ and $C_j$ with the same material parameters. The phase velocities need to be continuous because of the conservation of mass. We assume the phase pressures at the macroscale to be continuous as these represent the average of the microscopic phase pressures in an REV. \\
		Using these assumptions the phase velocity $u_d$ can be approximated with the values at $C_i$ and $f_{ij}$ or with the values at $C_j$ and $f_{ij}$, 
		\begin{flalign} \label{eq:darcySol_phaseVelAtFace1}
			u_{d,f_{ij}} &= - M_{d,i} K_{0,i} \frac{p_{d,f_{ij}} - p_{d,i}}{0.5 \Delta x_{i,d_{ij}}} ,  \\ \label{eq:darcySol_phaseVelAtFace2}
			u_{d,f_{ij}} &= - M_{d,j} K_{0,j} \frac{p_{d,j} - p_{d,f_{ij}}}{0.5 \Delta x_{j,d_{ij}}} . 
		\end{flalign}
		We use these two equations to calculate the phase pressure at the face $f_{ij}$
		\begin{flalign} \label{eq:darcySol_phasePressAtFace}
			p_{d,f_{ij}} &= \frac{M_{d,i} K_{0,i} \Delta x_{j,d_{ij}} p_{d,i} + M_{d,j} K_{0,j} \Delta x_{i,d_{ij}} p_{d,j}}{M_{d,i} K_{0,i} \Delta x_{j,d_{ij}} + M_{d,j} K_{0,j} \Delta x_{i,d_{ij}}} . 
		\end{flalign}
		Inserting the face pressure back in Eq. \ref{eq:darcySol_phaseVelAtFace1} or Eq. \ref{eq:darcySol_phaseVelAtFace2} leads to
		\begin{flalign} \label{darcySol_phaseVelHarmonicAverage}
			u_{d,f_{ij}} &= \frac{2 \Delta x_{ij} M_{d,i} K_{0,i} M_{d,j} K_{0,j}}{ M_{d,i} K_{0,i} \Delta x_{j,d_{ij}} + M_{d,j} K_{0,j} \Delta x_{i,d_{ij}}} \frac{p_{d,i} - p_{d,j}}{\Delta x_{ij}} = \overline{M_d K_0}^{H,f_{ij}}  \frac{p_{d,i} - p_{d,j}}{\Delta x_{ij}} .
		\end{flalign}
		This shows that the continuity of the velocity and phase pressure lead to the usage of the harmonic mean of $M_d K_0$ at faces. \\
		We discretize the respective terms in all velocities consistently with Eq. \ref{darcySol_phaseVelHarmonicAverage}, except for the total velocity $u$. The discrete formula for $u$ is given in Eq. \ref{discrete_totalVel}. For $u$, we use $\overline{M K_0}^{H,f_{ij}}$ as the coefficient of the non-wetting phase pressure gradient, and we refer to this discretization as the "Total Mobility" approach. The discretization consistent with Eq. \ref{darcySol_phaseVelHarmonicAverage} would instead use 
		\begin{flalign}
			\overline{M_w K_0}^{H,f_{ij}} + \overline{M_n K_0}^{H,f_{ij}}
		\end{flalign}		
		as the coefficient of the non-wetting phase pressure gradient. We refer to this as the "Phase Mobility" discretization, and the two coefficients are not equal. We accept the inconsistency of the "Total Mobility" discretization, because it leads to a better mass conservation for compressible fluid phases in the example of Section \ref{sec:darcySol_compressGas}. This comparison is presented in Table \ref{tab:darcySol_compressGasNWPMassVsIMPESIter}. \\

		In the discrete pressure equation, Eq. \ref{discretePressEq}, values of the phase densities at faces are needed. We calculate these values by calculating the phase pressures $p_w$ and $p_n$ at faces and use these to determine the respective phase densities. If the phase densities depend additionally on other variables like the temperature, also these variables must be calculated on the faces. \\
		Consistently with Eq. \ref{eq:darcySol_phasePressAtFace}, we determine the phase pressures at a face $f_{ij}$ by
		\begin{flalign} \label{eq:darcySol_facePhasePress}
			p_{d,f_{ij}}^{k_{l-1}} &= \frac{M_{d,i}^{k_{l}} K_{0,i} \Delta x_{j,d_{ij}} p_{d,i}^{k_{l-1}} + M_{d,j}^{k_{l}} K_{0,j} \Delta x_{i,d_{ij}} p_{d,j}^{k_{l-1}}}{M_{d,i}^{k_{l}} K_{0,i} \Delta x_{j,d_{ij}} + M_{d,j}^{k_{l}} K_{0,j} \Delta x_{i,d_{ij}}} , \\ \label{eq:darcySol_faceCapPress}
			p_{c,f_{ij}}^{k_{l-1}} &= p_{n,f_{ij}}^{k_{l-1}} - p_{w,f_{ij}}^{k_{l-1}} .
		\end{flalign}

		\subsubsection{Calculation of the Face Velocities} \label{sec_faceVel}
		In this section, Step \ref{alg_faceVelocities} of the solver is explained in more detail. \\
		The velocities $u$ and $u_D$ on faces are approximated by
		\begin{flalign} \label{discrete_totalVel}
			u^{k_l}_{f_{ij}} = \;  \text{sign} (n_{ij}) & \left( - \overline{M^{k_l} K_0 }^{H,f_{ij}} \nabla p^{k_l}_{n,f_{ij}} \cdot n_{ij} + \overline{M_w^{k_l} K_0 }^{H,f_{ij}} \nabla p^{k_l}_{c,f_{ij}} \cdot n_{ij}  \right) \\
			& \; + \left( \overline{ M_w^{k_l} K_0 }^{H,f_{ij}} \overline{ \rho_w^{k_l} }^{A,f_{ij}} + \overline{ M_n^{k_l} K_0 }^{H,f_{ij}} \overline{ \rho_n^{k_l} }^{A,f_{ij}} \right) g , \\ \label{eq:darcySol_discreteDarcyVel}
			u^{k_l}_{D,f_{ij}} = \; \text{sign} (n_{ij} & ) \overline{K_0 }^{H,f_{ij}}  \nabla p^{k_l}_{c,f_{ij}} \cdot n_{ij} + \overline{K_0 }^{H,f_{ij}} \left( \overline{ \rho_w^{k_l} }^{A,f_{ij}} - \overline{ \rho_n^{k_l} }^{A,f_{ij}} \right) g .
		\end{flalign}
		These velocities are scalar valued because only the component normal to the respective face is calculated.

		\subsubsection{Generalized Characteristic Wave Velocity Timestep Criterion} \label{sec_charWaveVel}
		
		In this section the timestep criterion used in Step \ref{alg_timestep} of the IMPES solver is explained. \\
		We refer to this timestep criterion in the following as \textbf{generalized characteristic velocity criterion} and it is a generalization of the timestep criterion explained for instance by \cite{lamine2015multidimensional}. \\
		
		The timestep criterion is based on the CFL condition \parencite{moura2012courant}. To derive it, we consider the saturation equation, Eq. \ref{two_phase_darcy_comp_2fluid_satEq}, and neglect the compressibility of the fluid phases,
		\begin{equation} \label{timestepSatEqIncomp}
			\frac{\partial S_w}{\partial t} = - \frac{1}{\phi} \nabla \cdot  u_w = - \frac{1}{\phi} \frac{\partial u_w}{\partial S_w} \cdot \nabla S_w.
		\end{equation}
		We assume that ignoring compressibility in the timestep criterion does not substantially impact the flow or lead to instabilities. In Section \ref{sec:darcySol_compressGas} the timestep criterion demonstrates its capability to stabilize two-phase flow with compressible fluid phases, despite the assumption of incompressibility in the derivation. \\
		
		If we assume for a moment that the derivative of $u_w$ in Eq. \ref{timestepSatEqIncomp} is constant, the CFL condition of this linear system reads
		\begin{flalign} \label{timeStepContLinear}
			\Delta t \leq \underset{C_i \in \mathbb{V}}{\text{min}} \left( \frac{ \phi_i \; C_{stab} }{ \sum^{n_D}_{d=1} \frac{1}{\Delta x_{i,d}} \left| \omega_{ij} \right| } \right) ,
		\end{flalign}
		where 
		\begin{flalign} \label{eq:theory_charWaveVel}
			\omega_{ij}(S_w) := \left. \frac{\partial u_w}{\partial S_w} \right|_{f_{ij}} \cdot n_{ij}  \overset{ \text{Eq. } \ref{u_w}}{=} \left. \left( \frac{\partial f_w}{\partial S_w}  u + \frac{\partial \gamma}{\partial S_w} u_D + f_w \frac{\partial u}{\partial S_w} + \gamma \frac{\partial u_D}{\partial S_w} \right)  \right|_{f_{ij}} \cdot n_{ij} ,
		\end{flalign}
		is the so-called characteristic wave velocity on face $f_{ij}$ and $C_{stab} > 0$ is the so-called stability constant. \\
		In the nonlinear case, the characteristic wave velocity $\omega_{ij}$ varies with saturation and is not constant. To obtain a computationally efficient CFL condition for the non-linear scenario, we use Eq. \ref{timeStepContLinear} while maximizing $\omega_{ij}$ over the local saturation values. \\

		The derivatives of $f_w$ and $\gamma$ are easy to compute, because in these terms only the relative permeabilities are dependent on the saturation. The derivatives of $u$ and $u_D$ are more complicated. Instead of an exact formula, we develop a numerical approximation of these derivatives. \\
		For the approximation of the derivative of $u$, we consider
		\begin{flalign} \label{chainRuleSpatial}
			n_{ij}^T \nabla u(S_w, p_n) n_{ij} = \left( \frac{\partial u}{\partial S_w} \cdot n_{ij} \right) \left( \nabla S_w \cdot n_{ij} \right) + \left( \frac{\partial u}{\partial p_n} \cdot n_{ij} \right) \left( \nabla p_n \cdot n_{ij} \right) .
		\end{flalign}
		We approximate the gradients of $u$ and $S_w$ on a face $f_{ij}$ by
		\begin{flalign*}
			\left. \nabla S_w \right|^{k-1}_{f_{ij}} \cdot n_{ij} &\approx  \frac{S^{k-1}_{w,i} - S^{k-1}_{w,j}} {\Delta x_{ij}} , \\
			n_{ij}^T \left. \nabla u \right|^{k-1}_{f_{ij}} n_{ij} &\approx  \frac{ \left( u^{k-1}_{i} - u^{k-1}_{j} \right) \cdot n_{ij} } {\Delta x_{ij}} .
		\end{flalign*}
		We substitute these approximations into Eq. \ref{chainRuleSpatial} and omit the term involving the gradient of $p_n$ on the right-hand side. Without this simplification, deriving a straightforward numerical approximation of the derivative of $u$ would not be feasible. Moreover, the approximation mainly aims to capture the capillary effects, where the omitted term plays a minor role. \\
		This results in
		\begin{flalign} \notag
			\frac{ \left( u^{k-1}_{i} - u^{k-1}_{j} \right) \cdot n_{ij} } {\Delta x_{ij}} &\approx \left( \left. \frac{\partial u}{\partial S_w} \right|_{f_{ij}} \cdot n_{ij} \right) \frac{S^{k-1}_{w,i} - S^{k-1}_{w,j}} {\Delta x_{ij}} .
		\end{flalign}
		If $S^{k-1}_{w,i} - S^{k-1}_{w,j} \neq 0$ this can be rearranged to an approximation of the derivative of $u$ with respect to the saturation $S_w$. \\
		If not, we can do the same procedure but with a time derivative in Eq. \ref{chainRuleSpatial} instead of a spatial derivative. In this case, we get an approximation, if $S^{k-1}_{w,f_{ij}} - S^{k-2}_{w,f_{ij}} \neq 0$. \\
		To summarize, we get the approximation 
		\begin{flalign} \label{velDerivSat}
			\left. \frac{\partial u}{\partial S_w} \right|^{k-1}_{f_{ij}}  \cdot n_{ij} &\approx D^{u, k-1}_{f_{ij}} := \left\{\begin{array}{ll} \frac{\left( u^{k-1}_{i} - u^{k-1}_j \right) \cdot n_{ij} }{S^{k-1}_{w,i} - S^{k-1}_{w,j}} , &\text{if} \; \; \left| S^{k-1}_{w,i} - S^{k-1}_{w,j} \right| \geq \Delta^{\text{min}}_s , \\
				\frac{ \left( u^{k-1}_{f_{ij}} - u^{k-2}_{f_{ij}} \right) \cdot n_{ij} }{S^{k-1}_{w,f_{ij}} - S^{k-2}_{w,f_{ij}}} , \; \;  &\text{else if } \left| S^{k-1}_{w,f_{ij}} - S^{k-2}_{w,f_{ij}} \right| \geq \Delta^{\text{min}}_t , \\
				0, \; \; &\text{else, } \end{array} \right.
		\end{flalign}
		where $\Delta^{\text{min}}_s > 0$ and $\Delta^{\text{min}}_t > 0$ are tolerances that ensure the denominators are not too small. \\
		The same approximation can also be done with $u_D$ instead of $u$. In that case, we denote the approximation by $D^{u_D, k-1}_{f_{ij}}$. \\
		
		Now, we maximize the characteristic wave velocity, Eq. \ref{eq:theory_charWaveVel}, on a face $f_{ij}$ in the form 
		\begin{flalign} \label{eq:darcySol_upperBoundCharWaveVel}
			\omega^{k,max}_{ij} &:= \underset{\hat{S}_w \in I^k_{ij}}{\text{max}} \left( \left| \frac{\partial f_w}{\partial S_w} \left( \hat{S}_w \right) u^{k-1}_{f_{ij}} + \frac{\partial \gamma}{\partial S_w} \left( \hat{S}_w \right) u^{k-1}_{D,f_{ij}} + f_w \left( \hat{S}_w \right) D^{u, k-1}_{f_{ij}} + \gamma \left( \hat{S}_w \right) D^{u_D, k-1}_{f_{ij}}  \right|  \right) , \\ \notag
			&\approx \underset{\hat{S}_w \in I^k_{ij}}{\text{max}}  \left( \left| \omega_{ij} \left( \hat{S}_w \right)  \right|  \right)
		\end{flalign}
		where the interval of the maximization is given by
		\begin{flalign} \label{intervalMax}
			I^k_{ij} = [S_{w,i}^{k-1}, S_{w,j}^{k-1}] \cup [S_{w,j}^{k-1}, S_{w,i}^{k-1}] .
		\end{flalign}
		
		Using this approximation in Eq. \ref{timeStepContLinear} leads to the timestep restriction
		\begin{flalign} \label{timeStepDiscr}
			\Delta t^{gcw}_k := \underset{C_i \in \mathbb{V}}{\text{min}} \left( \frac{ \phi_i \; C_{CFL} }{ \sum^{n_D}_{d=1} \frac{1}{\Delta x_{i,d}} \underset{f_{ij} \in F_{C_i}^{d}}{\text{max}} \omega^{k,max}_{ij} } \right) .
		\end{flalign}
		In this formula, we also maximize with respect to $F_{C_i}^{d}$. This set includes the two faces of $C_i$ that are orthogonal to the coordinate axis of space dimension $d$. \\

		The computational effort required by the timestep criterion is influenced by the performance of the scalar optimization method employed in Eq. \ref{eq:darcySol_upperBoundCharWaveVel}, since this maximization needs, in principle, to be done at every face of the grid. \\
		But, in many simulations, a large number of faces have adjacent cells with nearly identical saturations. For these cases, the optimization step can be omitted by directly using the maximum value calculated from the two saturations of the adjacent cells. In our approach, this shortcut is applied whenever the saturation difference of the two cells is below $10^{-6}$. When the difference is greater, we utilize an active set algorithm implementation \cite{bleicImplAlglib} to carry out the scalar maximization. \\
		We assume that there are no saturations for which both relative permeabilities are zero, and that the derivatives of both relative permeabilities are bounded for $S_w \in [0, 1]$. In Section \ref{section_capPressAndRelPermeab} the assumption of bounded derivatives is discussed for the Brooks-Corey relative permeabilities that are used in the examples in Section \ref{sec:numericalExperiments}. Under these two assumptions, the maximization in Eq. \ref{eq:darcySol_upperBoundCharWaveVel} is well-defined, as it involves maximizing a finite objective function over a finite interval. Moreover, the maximization algorithm cannot diverge, but it may converge to a local maximum rather than the global maximum. To reduce this risk, we perform the maximization starting from three different initial values, these are $S_{w,i}^{k-1}$,  $S_{w,j}^{k-1}$ and $\frac{1}{2} \left( S_{w,i}^{k-1} + S_{w,j}^{k-1} \right) $. \\
		As mentioned before, we use an active set algorithm \cite{harkegard2002efficient, best1990active} to solve the maximization problem. The stopping criterion is that the algorithm terminates when the change in the variable $\hat{S}_w$ between iterations is less than $10^{-6}$. Additionally, we set the maximum number of allowed iterations to $1000$. We note that this limit was never reached in the simulations presented in Section \ref{sec:numericalExperiments}. \\

		In the numerical experiments in Section \ref{sec:numericalExperiments}, we compare the effectiveness and reliability of the newly developed timestep criterion to already existing timestep criteria for IMPES solver in the literature. We compare it to the \textbf{Coats criterion} \parencite{coats2003impes} and the criterion explained by \cite{lamine2015multidimensional} to that we refer as \textbf{characteristic velocity criterion}. \\
		The timestep criterion given by Eq. \ref{timeStepDiscr} is based on a similar idea as the the characteristic velocity criterion. The difference is that in the characteristic velocity criterion the terms involving the derivatives of the velocities $u$ and $u_D$ in Eq. \ref{eq:theory_charWaveVel} are neglected. Accordingly, the numerical approximations of these velocity derivatives are not needed. \\
		The velocities $u$ and $u_D$ are dependent on the saturation through the relative permeabilities and the capillary pressure function. As the capillary pressure function largely varies with the saturation, it is to expect that these two criteria lead to significantly different timestep sizes, if capillary effects are important for the flow. \\
		
		The Coats criterion is derived from Neumann stability analysis. For two-phase flow the criterion is given by 
		\begin{flalign*}
			\Delta t_k^{Coats} &:= \underset{C_i \in \mathbb{V}}{\text{min}} \left( \frac{\phi_i |C_i| C_{\text{stab}}}{F^{k}_{i, Coats}} \right) , \\
			F^{k}_{i, Coats} &:= \sum_{f_{ij} \in F_{C_i}} \left| \theta^{k}_{ij, Coats} \right| , \\
			\theta^{k}_{ij, Coats} &:= |f_{ij}| \left( A^{k-1}_{f_{ij},Coats} \left| u^{k-1}_{w,f_{ij}} \right| + B^{k-1}_{f_{ij}, Coats} \left| u^{k-1}_{n,f_{ij}} \right| + C^{k-1}_{f_{ij},Coats} \left( \frac{\partial p^{k-1}_{C,i}}{\partial S_w} +  \frac{\partial p^{k-1}_{C,j}}{\partial S_w} \right) \right) .
		\end{flalign*}
		The coefficients are defined as	
		\begin{flalign*}
			A_{Coats} &:= \frac{M_{n}}{M M_{w}} \frac{\partial M_{w}}{\partial S_w} , \\
			B_{Coats} &:= \frac{M_{w}}{M M_{n}} \frac{\partial M_{n}}{\partial S_n} , \\
			C_{Coats} &:= - \frac{K_0}{d_{ij}} \frac{M_{n} M_{w}}{M} .
		\end{flalign*}
		The coefficients are computed at timestep $k-1$ on face $f_{ij}$ using upwinding. This upwinding is based on the non-wetting phase velocity $u^{k-1}_n$ when $M^{k-1}_n \neq 0$. If $M^{k-1}_n = 0$, the wetting phase velocity $u^{k-1}_w$ is used instead. \\

		Alongside any of the timestep criteria, we enforce a limit on the maximum relative increase of the timestep size in one time iteration. The parameter $\tau_{\text{max}} > 0$ sets this maximum allowed relative increase. This constraint prevents the solver from enlarging the timestep too abruptly, which might lead to instabilities.\\
		If $\Delta t^{\text{stab}}_k$ is the timestep size recommended by the criterion and $\Delta t_{k-1}$ is the previous timestep size, we select 
		\begin{flalign} \label{eq:theory_timestepCriterion}
			\Delta t_k := \text{min} \left\{ \Delta t^{\text{stab}}_k, \left( \tau_{\text{max}} + 1 \right)  \Delta t_{k-1} \right\}
		\end{flalign}
		as the final timestep size.

	\subsubsection{Typical Boundary Conditions} \label{sec_boundaryConditions}
	
		In this section, we present the boundary conditions used with the two-phase Darcy equations. The velocity is uniquely determined by saturation and pressure. Therefore, boundary conditions for velocity are generally unnecessary, but in the case of wall boundaries, we prescribe the fluxes to be zero across the boundary.  \\
		
		Let $f_{ij}$ be a outer boundary next to the cell $C_{i}$. For simplicity, we introduce an imaginary cell $C_{j}$ outside the grid, positioned on the opposite side of the face $f_{ij}$, that has the same size as $C_{i}$. \\ 
		
		At \textbf{wall} boundaries, we set
		\begin{flalign*}
			u_{f_{ij}} = u_{D,f_{ij}} &= u_{w,f_{ij}} = u_{n,f_{ij}} = 0 .
		\end{flalign*}	
		As \textbf{symmetric} boundary conditions, we indicate
		\begin{flalign*}
			\nabla p_{n, f_{ij}} \cdot n_{ij} &= 0 , \\
			\nabla S_{w, f_{ij}} \cdot n_{ij} &= 0 .
		\end{flalign*}
		At an \textbf{outlet} boundary, we prescribe 
		\begin{flalign*}
			\nabla S_{w, f_{ij}} \cdot n_{ij} &= 0
		\end{flalign*}
		and a Dirichlet boundary condition for the pressure $p_n$. In most cases the boundary value is $10^5 Pa$ that is the atmospheric pressure. \\
		
		At an \textbf{inlet} boundary, we set 
		\begin{flalign*}
			S_{w, f_{ij}} &= 1
		\end{flalign*}
		as saturation boundary condition. In most experiments, we also apply a Dirichlet boundary condition for the pressure, setting the boundary pressure value equal to or higher than that at the inlet boundary. \\		
		In the example of Section \ref{sec:darcySol_BL} that is not the case. There we prescribe a constant inflow velocity $u_{In} > 0$ at the inlet. Because gravity and capillary forces are neglected in this example, it is 
		\begin{flalign} \label{eq:darcySol_BLbndCond}
			u_{In} = M K_0 \nabla p_{n, f_{ij}} \; \; \Leftrightarrow \; \; \nabla p_{n, f_{ij}}  = \mu_w K_0^{-1} u_{In} .
		\end{flalign}
		Equation \ref{eq:darcySol_BLbndCond} implements the constant inflow velocity condition through a Neumann boundary condition on the pressure.

	\section{Results and Discussion} \label{sec:numericalExperiments}
	
		In the following sections, we provide numerical results for testing and validating the two-phase Darcy solver. The examples include scenarios involving pressure drop dominated flow, capillary dominated flow, flow with compressible fluid phases, and flow featuring discontinuous material parameters. Additionally, we compare the newly introduced timestep criterion from Section \ref{sec_charWaveVel} with both the Coats criterion and the characteristic wave velocity criterion. \\
		Unless otherwise specified, all subsequent simulations employ the generalized characteristic wave velocity criterion together with the relative increase restriction on the timestep size from Eq. \ref{eq:theory_timestepCriterion}. The parameters are set to $\tau_{\text{max}} = 0.01$, $C_{\text{stab}} = 1$, and $\Delta^{\text{min}}_s = \Delta^{\text{min}}_t = 10^{-4}$, unless otherwise indicated.

		\subsection{Capillary Rise} \label{sec:darcySol_simpleCapRise}
		
		The first example is a capillary rise experiment. The goal of this example is to assess the effectiveness of the timestep criteria in stabilizing capillary dominated flows. \\
		
		\begin{table}[ht]
			\centering
			\begin{tabular}{@{}c|c|c|c|c|c@{}}
				$\phi$ & $K_0$  &  $\mu_w$ &  $\mu_n$ &  $\rho_w$ &  $\rho_n$  \\
				\hline
				$0.4$ & $5.8 \cdot 10^{-13} m^2$ & $ 6.72 \cdot 10^{-2} Pa s $ & $1.76 \cdot 10^{-5} Pa s $ & $ 920 \frac{kg}{m^3}$ & $ 1.22 \frac{kg}{m^3}$   \\
			\end{tabular}
			\caption{ Both fluid phases are assumed to be incompressible. } 
			\label{tab:darcySol_paramsSimpleCapRise}
		\end{table}
		
		\begin{figure}[ht]
			\centering
			\includegraphics[width=0.6\textwidth]{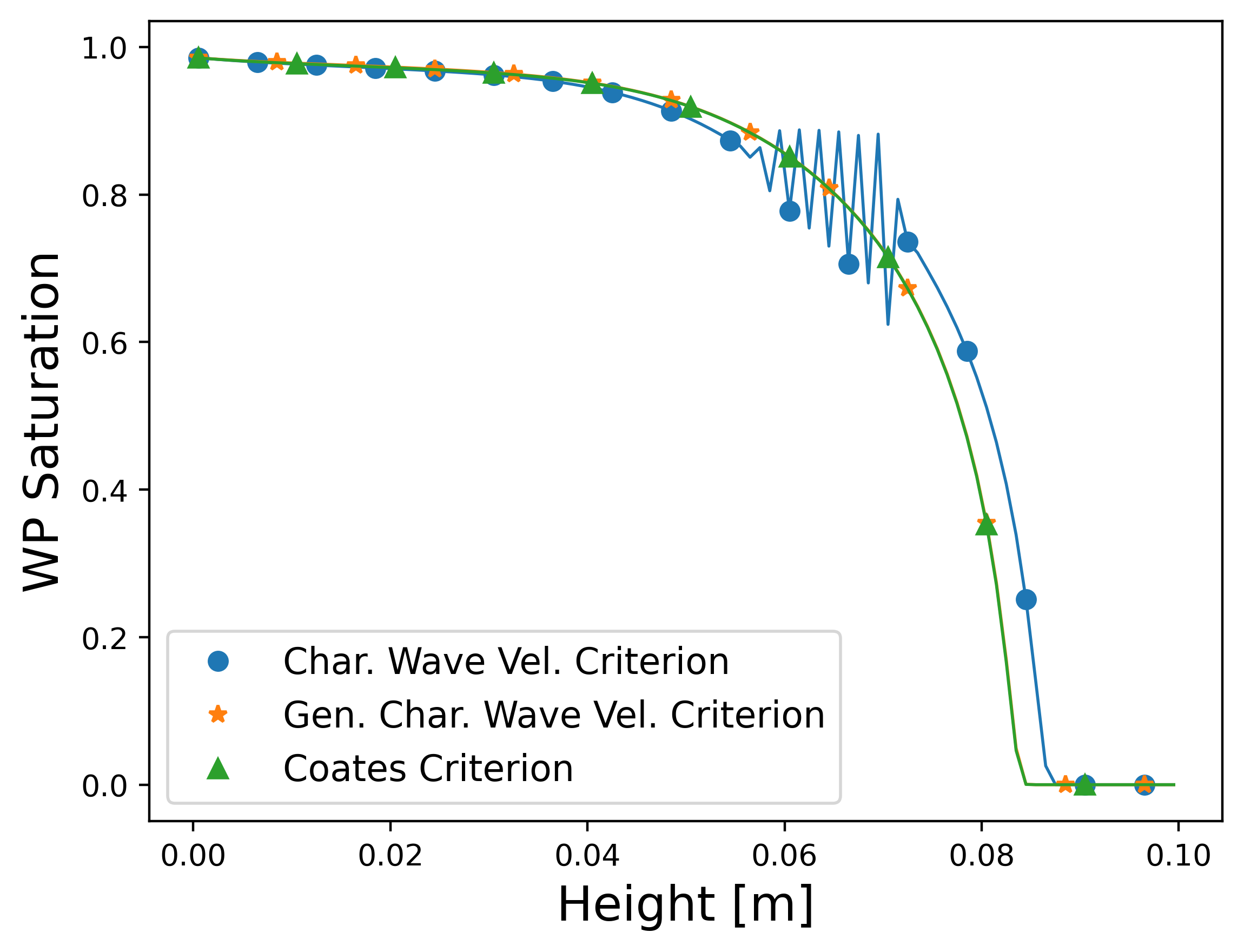}
			\caption{ The image displays the saturation profiles from the capillary rise simulations at $10000 s$. The results were obtained using the characteristic wave velocity criterion, the generalized characteristic wave velocity criterion, and the Coats criterion. The grid used measures $10 \times 100$ voxels. }
			\label{fig:darcySol_simpleCapRiseSatVsHeight}
		\end{figure}
		
		We simulate the capillary rise in a $0.1 m$ high column that is initially filled with air. Gravity is directed in the negative height direction and the wetting phase enters the column from the bottom. Therefore, the bottom boundary serves as an inlet, while the top boundary is an outlet. We set the same Dirichlet pressure values at in- and outlet. The physical parameters are summarized in Table \ref{tab:darcySol_paramsSimpleCapRise}. The non-wetting phase is air and the wetting phase correspond to a light polymer resin. Relative permeabilities are described using the simplified Brooks-Corey functions with $m^{w,BC} = m^{nw,BC} = 4$. The capillary pressure is a Van Genuchten function with parameters $p^{VG}_e = 17.7 k Pa$ and $m^{VG} = 0.74$. \\ 
		The simulation is performed with $C_{\text{stab}} = 1$ applying all three timestep criteria. The computational domain uses a two-dimensional grid with $100$ cells in height and $10$ cells in width. Symmetric boundary conditions are set on the left and right boundaries of the geometry. Since the setup is uniform in the width direction, results are presented as a one-dimensional profile along the height. \\
		Figure \ref{fig:darcySol_simpleCapRiseSatVsHeight} shows the saturation profiles along the height after $10000 s$ of all three simulations. The characteristic wave velocity criterion exhibits significant oscillations, whereas the results from both the generalized characteristic wave velocity criterion and Coats criterion are nearly identical and free of oscillations. \\
		The results clearly highlight the limitations of the characteristic wave velocity criterion in stabilizing flows dominated by capillarity, and the results confirm that the additional terms in the generalized characteristic wave velocity criterion enhances its ability the stabilize capillary dominated flows.

		\subsection{Buckley-Leverett Problem} \label{sec:darcySol_BL}
		
		To validate the IMPES solver for pressure drop dominated flows, we apply it to the Buckley-Leverett problem. This problem considers one-dimensional flow driven by a pressure drop. Capillary pressure effects and gravity are neglected. As demonstrated by \cite{buckley1942mechanism}, an analytic solution exists that can be calculated. Therefore, the Buckley-Leverett problem is well suited for validating numerical two-phase flow solvers in the case of pressure drop dominated flows. \\
		
		\begin{table}[ht]
			\centering
			\begin{tabular}{@{}c|c|c|c@{}}
				$\phi$ & $K_0$  &  $\mu_w$ &  $\mu_n$   \\
				\hline
				$0.4$ & $5 \cdot 10^{-13} m^2$ & $1 \cdot 10^{-4} Pa s $ & $1 \cdot 10^{-4} Pa s $   \\
			\end{tabular}
			\caption{ Both fluid phases are assumed to be incompressible. } 
			\label{tab:darcySol_paramsBL}
		\end{table}

		The experiment uses a one-meter-long geometry with an inlet at the left end and an outlet at the right end. Initially, the geometry is filled with the non-wetting phase. We choose the viscosities of both phases to be equal, then the saturation at the shock front in the analytic solution is significantly less than one. In this case, the rarefaction wave part of the analytic solution is clearly visible. This would not be the case, if the viscosity of the non-wetting phase were much smaller than that of the wetting phase. A constant inflow velocity of $2.5 \cdot 10^{-4} \frac{m}{s}$ is imposed at the inlet as in Eq. \ref{eq:darcySol_BLbndCond}. The physical parameters are listed in Table \ref{tab:darcySol_paramsBL}. Moreover, we use simplified Brooks-Corey relative permeabilities with $m^{w,BC} = m^{nw,BC} = 4$. \\
		Similar to the previous example, the simulations are performed on a two-dimensional grid with symmetric boundary conditions on the additional boundaries. Since the setup and results are uniform in this direction, only the results along the non-constant direction are presented. \\

		\begin{figure}[ht]
			\centering
			\includegraphics[width=0.6\textwidth]{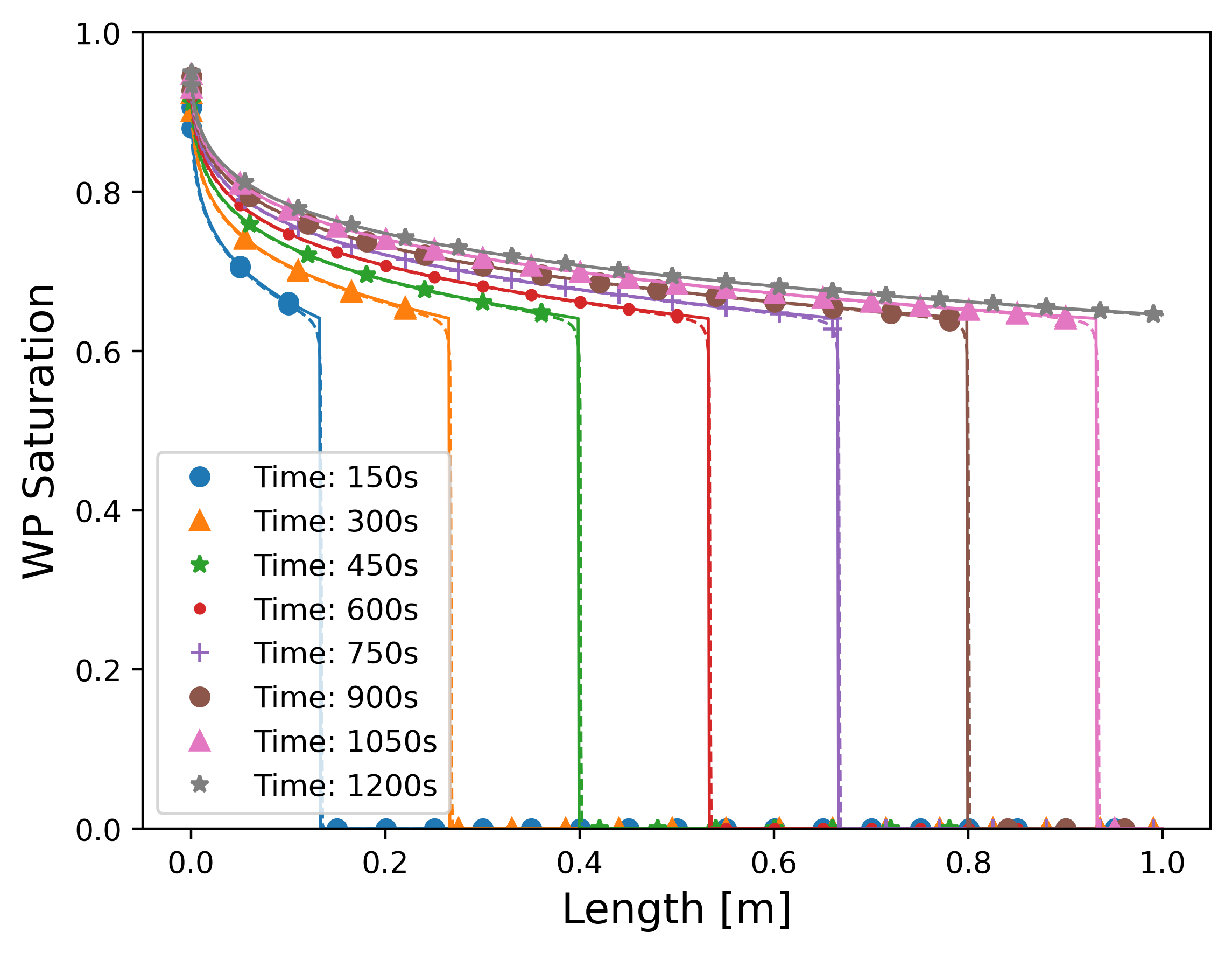}
			\caption{ The numerical and analytic solutions of the Buckley-Leverett problem at various times are shown. Solid lines correspond to the analytic solution, while dashed lines indicate the numerical results. The geometry is discretized with a voxel size of $10^{-3} m$, and the simulations are performed using $C_{\text{stab}} = 1$. }
			\label{fig:darcySol_BLCompNumToAnalySol}
		\end{figure}
		
		\begin{figure}[ht]
			\centering
			\includegraphics[width=0.6\textwidth]{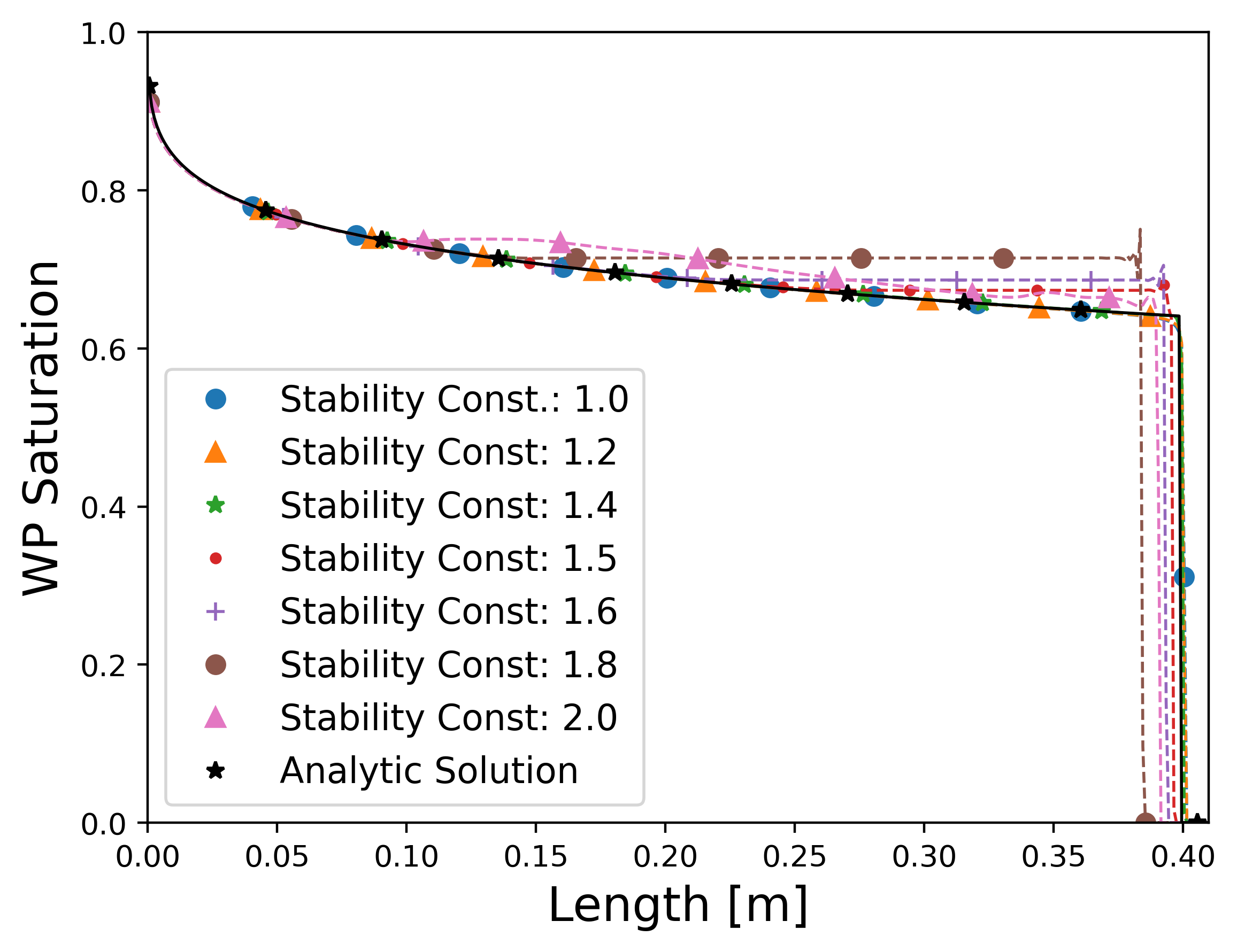}
			\caption{ This image shows both the analytic and numerical solutions of the Buckley-Leverett problem at time $t=450 s$. The geometry is discretized with a voxel size of $10^{-3} m$ and the numerical solutions are computed using various stability constants $C_{\text{stab}}$. }
			\label{fig:darcySol_BLOscialltions}
		\end{figure}
		
		\begin{figure}[ht]
			\begin{subfigure}[b]{0.5\textwidth}
				\centering
				\includegraphics[width=\textwidth]{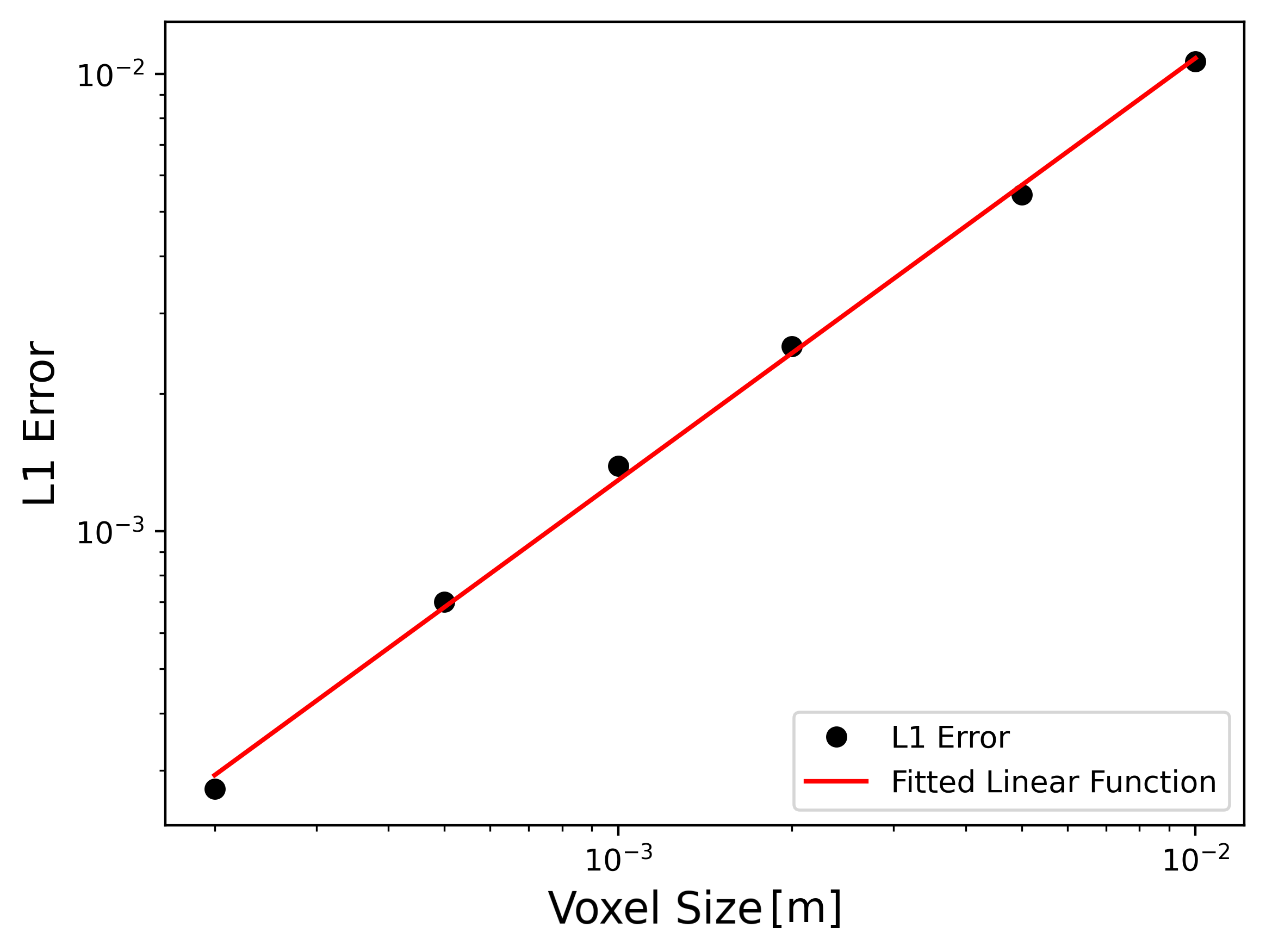}
			\end{subfigure}
			\begin{subfigure}[b]{0.5\textwidth}
				\centering
				\includegraphics[width=\textwidth]{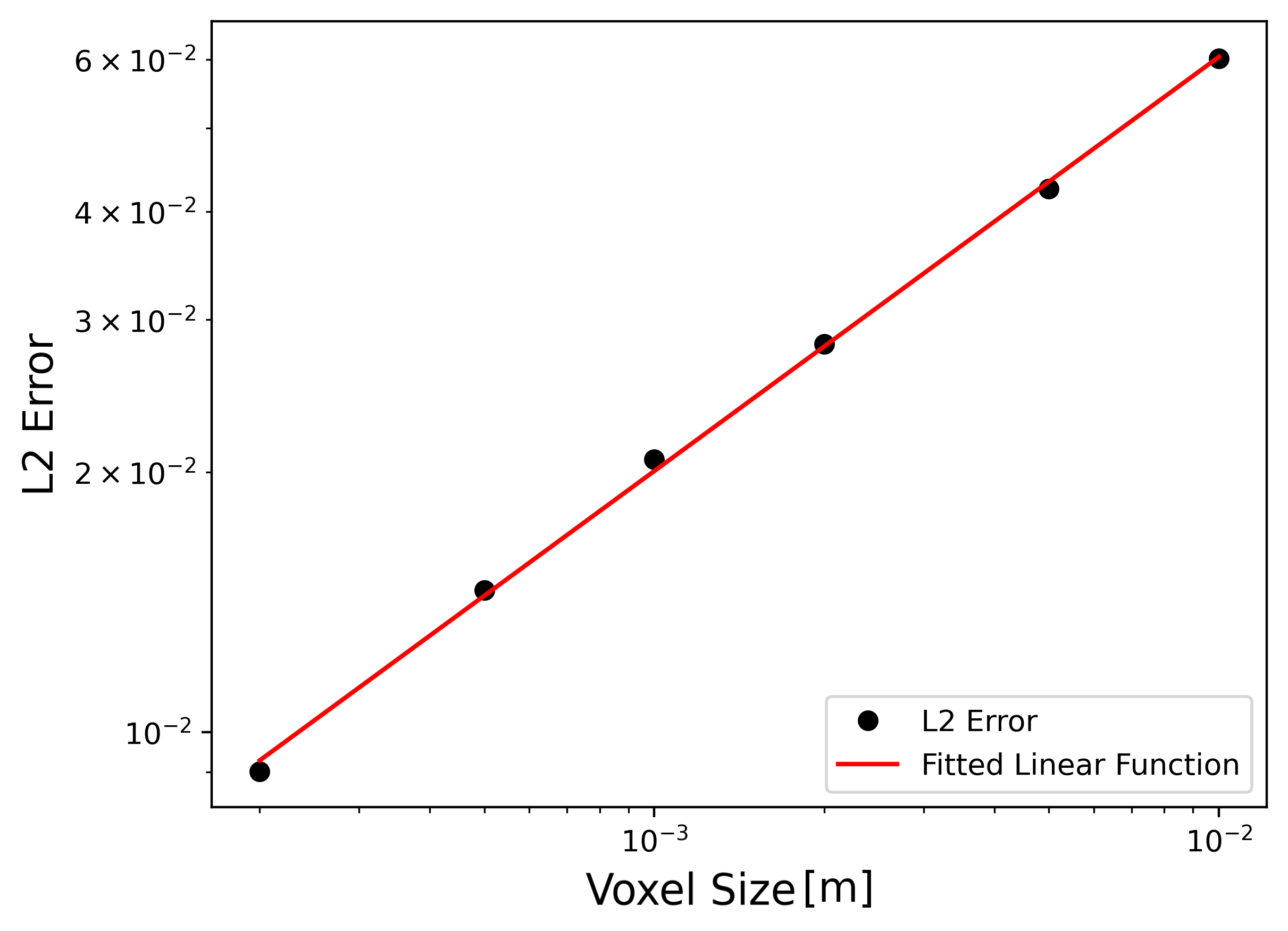}
			\end{subfigure}
			\caption{ These two images display the errors of the numerical solutions for the Buckley-Leverett problem across different voxel sizes. The stability constant is set to $C_{\text{stab}} = 1$. Black dots indicate the numerical errors, with the left image presenting the $L_1$ error and the right image showing the $L_2$ error. The red lines correspond to linear least-square regression lines, estimating the order of convergence of the solver. These regression lines are computed by a least-squares fit of the ten-logarithm of the voxel sizes and the ten-logarithm of the corresponding error values. The slope of this fit represents the approximate order of convergence, yielding about $0.92$ for the $L_1$ error and $0.48$ for the $L_2$ error. }
			\label{fig:darcySol_convPlotBL}
		\end{figure}

		Figure \ref{fig:darcySol_BLCompNumToAnalySol} presents both the analytic and numerical solutions of the Buckley-Leverett problem at various times. The analytic solution consists of a shock wave followed by a rarefaction wave and the numerical solution \ref{fig:darcySol_BLCompNumToAnalySol} demonstrates that the presented IMPES solver closely matches the analytic solution. The greatest numerical error appearing right at the shock front. \\ \\
		In Figure \ref{fig:darcySol_BLOscialltions}, the analytic solution at time $t=450s$ is compared with numerical solutions using stability constants between $1$ and $2$. The figure shows that numerical results remain stable up to a stability constant of $1.4$, while higher values lead to oscillations near the shock front. \\
		
		Figure \ref{fig:darcySol_convPlotBL} displays the $L_1$ and $L_2$ errors of the numerical solutions over different voxel sizes. For this example, errors are calculated by comparing wetting phase saturation values to the analytic solution at $101$ equidistant time steps between $t=0$ and $t=1500 s$. The maximum difference across all time steps is reported as the overall error of the numerical solution. The order of convergence is estimated by a linear least-squares fit, yielding about $0.92$ for the $L_1$ error and $0.48$ for the $L_2$ error. \\ \\

		\begin{table}[ht]
			\centering
			\caption*{ }
			\begin{tabular}{@{}c|c|c|c|c|c|c|c|c@{}}
				$C_{\text{stab}}$ & $ 1.0 $ & $ 1.1 $ & $ 1.2 $ & $ 1.3 $ & $ 1.4 $ & $ 1.5 $ & $ 1.6 $ & $ 1.7 $ \\
				\hline
				$L_1$ Error in $[10^{-3}]$ & $ 1.33 $ & $ 1.01 $ & $ 1.03 $ & $ 1.01 $ & $ 0.78 $ & $ 0.78 $ & $ 673 $ & $ 646 $  \\
				\hline
				$L_2$ Error in $[10^{-2}]$ & $ 1.97 $ & $ 1.80 $ & $ 1.87 $ & $ 1.97 $ & $ 1.53 $ & $ 1.43 $ & $ 69.1 $ & $ 67.2 $  \\
				\hline
				Total Iter. Nb. & $ 4076 $ & $ 3389 $ & $ 3179 $ & $ 2900 $ & $ 2424 $ & $ 2396 $ & $ 82 $ & $ 128 $   \\
				\hline
				Average $ \Delta t$ in $[s]$ & $ 0.37 $ & $ 0.44 $ & $ 0.47 $ & $ 0.52 $ & $ 0.62 $ & $ 0.63 $ & $ 18.32 $ & $ 11.96 $
			\end{tabular}
			\caption*{\textbf{Coats Criterion}}

			\begin{tabular}{@{}c|c|c|c|c|c|c|c|c@{}}
				$C_{\text{stab}}$ & $ 1.0 $ & $ 1.1 $ & $ 1.2 $ & $ 1.3 $ & $ 1.4 $ & $ 1.5 $ & $ 1.6 $ & $ 1.7 $   \\
				\hline
				$L_1$ Error in $[10^{-3}]$ & $ 1.28 $ & $ 1.23 $ & $ 1.17 $ & $ 1.07 $ & $ 0.80 $ & $ 12.3 $ & $ 22.4 $ & $ 35.4 $  \\
				\hline
				$L_2$ Error in $[10^{-2}]$ & $ 1.99 $ & $ 1.97 $ & $ 1.99 $ & $ 2.03 $ & $ 1.67 $ & $ 6.24 $ & $ 8.58 $ & $ 10.9 $    \\
				\hline
				Total Iter. Nb. & $ 3414 $ & $ 3105 $ & $ 2847 $ & $ 2628 $ & $ 2466 $ & $ 2186 $ & $ 2015 $ & $ 1873 $   \\
				\hline
				Average $ \Delta t$ in $[s]$ & $ 0.44 $ & $ 0.48 $ & $ 0.53 $ & $ 0.57 $ & $ 0.61 $ & $ 0.69 $ & $ 0.75 $ & $ 0.80 $   
			\end{tabular}
			\caption*{\textbf{Characteristic Wave Velocity Criterion}}
			
			\begin{tabular}{@{}c|c|c|c|c|c|c|c|c@{}}
				$C_{\text{stab}}$ & $ 1.0 $ & $ 1.1 $ & $ 1.2 $ & $ 1.3 $ & $ 1.4 $ & $ 1.5 $ & $ 1.6 $ & $ 1.7 $  \\
				\hline
				$L_1$ Error in $[10^{-3}]$ & $ 1.28 $ & $ 1.23 $ & $ 1.17 $ & $ 1.06 $ & $ 0.83 $ & $ 12.3 $ & $ 22.4 $ & $ 35.4 $    \\
				\hline
				$L_2$ Error in $[10^{-2}]$ & $ 1.99 $ & $ 1.97 $ & $ 1.99 $ & $ 2.03 $ & $ 1.64 $ & $ 6.24 $ & $ 8.58 $ & $ 10.9 $   \\
				\hline
				Total Iter. Nb. & $ 3422 $ & $ 3114 $ & $ 2855 $ & $ 2635 $ & $ 2476 $ & $ 2214 $ & $ 2063 $ & $ 1948 $   \\
				\hline
				Average $ \Delta t$ in $[s]$ & $ 0.44 $ & $ 0.48 $ & $ 0.53 $ & $ 0.57 $ & $ 0.61 $ & $ 0.68 $ & $ 0.73 $ & $ 0.77 $
			\end{tabular}
			\caption*{\textbf{Generalized Characteristic Wave Velocity Criterion}}
			
			\caption{ \label{tab:darcySol_BLTimestepComparison} These tables list the errors in the $L_1$ norm, the errors in the $L_2$ norm, the total number of time iterations, and the arithmetic mean of all timestep sizes for simulations of the Buckley-Leverett problem. The first table uses the Coats criterion, the second table uses the characteristic wave velocity criterion, and the last table uses the generalized characteristic wave velocity criterion. In each case, a range of values for the stability constant $C_{\text{stab}}$ is tested. All simulations were performed with $\tau_{\text{max}} = 0.3$, an initial timestep size of $1 \cdot 10^{-2} s$, and a voxel size of $10^{-3} m$. } 
		\end{table}

		To compare the timestep criteria, the setup is simulated using all three timestep criteria with various stability constants $C_{\text{stab}}$. The Table \ref{tab:darcySol_BLTimestepComparison} shows error and timestep statistics from these simulations. In these simulations $\tau_{\text{max}} = 0.3$ is used. This allows the timestep size to increase up to a $30 \%$ between consecutive time iterations to reduce the impact of this parameter on the comparison. \\
		To measure the computational effort, the tables list the total number of time iterations and the average timestep size, calculated as the arithmetic mean of all timestep sizes. Accuracy is assessed through the $L_1$ and $L_2$ norms measuring the difference from the analytic solution.
		A comparison of the $L_1$ and $L_2$ errors in Table \ref{tab:darcySol_BLTimestepComparison} of the generalized characteristic wave velocity criterion with the saturation profiles in Figure \ref{fig:darcySol_BLOscialltions} reveals that the increase in errors aligns with the onset of oscillations for $C_{\text{stab}} > 1.4$. Both characteristic wave velocity criteria maintain stability up to $C_{\text{stab}} = 1.4$, whereas the Coats criterion remains stable up to $C_{\text{stab}} = 1.5$. The two characteristic wave velocity criteria behave similarly on this example. The absence of capillary effects leads likely to relatively small saturation derivatives of velocities $u$ and $u_D$ and then these two criteria are similar. \\
		At $C_{\text{stab}} = 1.0$, the Coats criterion requires about $19 \%$ less time iterations than the generalized characteristic wave velocity criterion. When comparing each criterion at its maximum stable value of $C_{\text{stab}}$ , the difference in total iterations is minimal. However, identifying this precise stability constant for which a timestep criterion is barely stable is usually impractical for larger and more complex applications. Moreover, such stability constants are typically chosen with some "safety distance". \\
		
		In summary, the introduced solver accurately captures the pressure drop dominated flow of the Buckley-Leverett problem. Additionally, the Coats criterion needs about $19 \%$ more time iterations than the newly proposed timestep criterion, while maintaining a similar accuracy. Moreover, it needs a similar number of time iterations as the characteristic wave velocity criterion. \\
		Using the same hardware and number of threads, the simulation with the generalized characteristic wave velocity criterion and $C_{\text{stab}} = 1$ takes $193.87s$, whereas the simulation with the Coats criterion takes $229.57s$. Thus, the generalized characteristic wave velocity criterion requires approximately $16 \%$ less runtime than the Coats criterion. This demonstrates that the generalized characteristic wave velocity criterion is not significantly more computationally complex than the Coats criterion.

		\subsection{Capillary Gravity Equalization}  \label{sec:darcySol_capGravEq}
		
		In this section, we assess the IMPES solver using an example where flow is driven by capillary pressure and gravity. This example was also used by \cite{horgue2015open} to test a numerical solver. \\
		The geometry is two-dimensional, measuring $0.1 m$ in width and $1 m$ in height, and filled with a homogeneous porous medium. On the bottom, right, and left boundaries, we impose wall boundary conditions. The bottom, right, and left boundaries are walls, i.e., there can be no flow across them. At the top boundary, Dirichlet conditions are applied for both pressure and saturation, with pressure fixed at $10^{5} Pa$ (atmospheric pressure) and saturation set to $10^{-6}$. \\
		Initially, the saturation is $0.5$ in the lower half and $10^{-6}$ in the upper half of the geometry. The gravity is directed in the negative height direction. \\
		
		\begin{table}[ht]
			\centering
			\begin{tabular}{@{}c|c|c|c|c|c@{}}
				$\phi$ & $K_0$  &  $\mu_w$ &  $\mu_n$ &  $\rho_w$ &  $\rho_n$  \\
				\hline
				$0.5$ & $1 \cdot 10^{-11} m^2$ & $ 6.72 \cdot 10^{-2} Pa s $ & $1.76 \cdot 10^{-5} Pa s $ & $ 920 \frac{kg}{m^3}$ & $ 1.22 \frac{kg}{m^3}$   \\
			\end{tabular}
			\caption{ Both fluid phases are assumed to be incompressible. } 
			\label{tab:darcySol_paramsCapGravEq}
		\end{table}

		The simulations employ the physical parameters listed in Table \ref{tab:darcySol_paramsCapGravEq} and simplified Brooks-Corey relative permeabilities with $m^{w,BC} = m^{nw,BC} = 2$. The non-wetting phase is air and the wetting phase correspond to a light polymer resin. We consider two simulation cases. The first uses a Van Genuchten capillary pressure function with parameters $p^{VG}_e = 100 Pa$ and $m^{VG} = 0.5$, while the second uses a Brooks-Corey capillary pressure function with $p^{BC}_e = 1000 Pa$ and $m^{BC} = 2$. \\
		
		\begin{figure}[ht]
			\centering
			\includegraphics[width=0.6\textwidth]{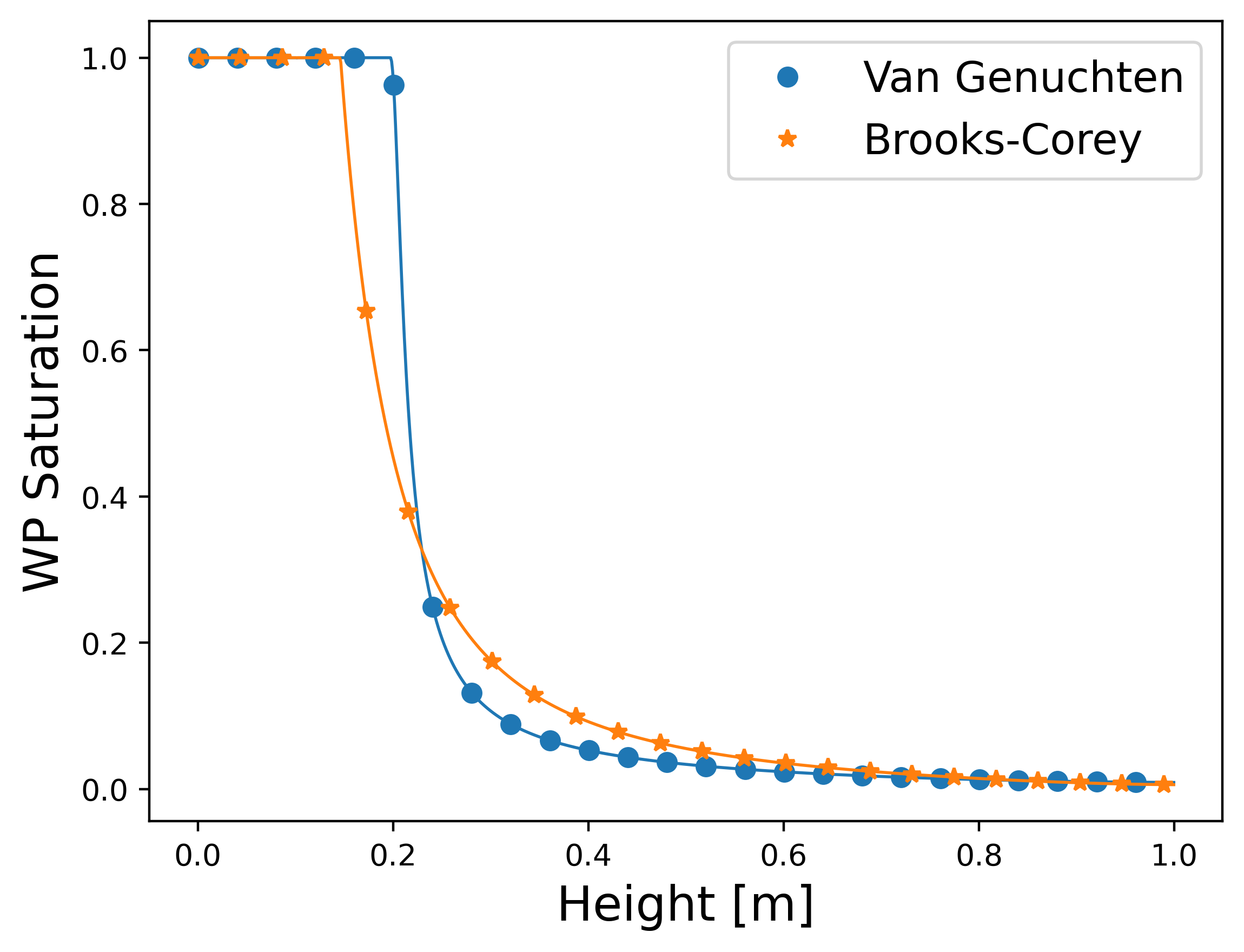}
			\caption{ This image depicts the stationary saturation fields from the numerical solutions. "Van Genuchten" refers to the use of the Van Genuchten capillary pressure function with parameters $p^{VG}_e = 100 Pa$ and $m^{VG} = 0.5$, while "Brooks-Corey" indicates the Brooks-Corey capillary pressure function with $p^{BC}_e = 1000 Pa$ and $m^{BC} = 2$. The grid used for the simulations consists of $10 \times 1000$ voxels. }
			\label{fig:darcySol_capGravEqSatVsHeight}
		\end{figure}
		\begin{figure}[ht]
			\begin{subfigure}[b]{0.5\textwidth}
				\centering
				\includegraphics[width=\textwidth]{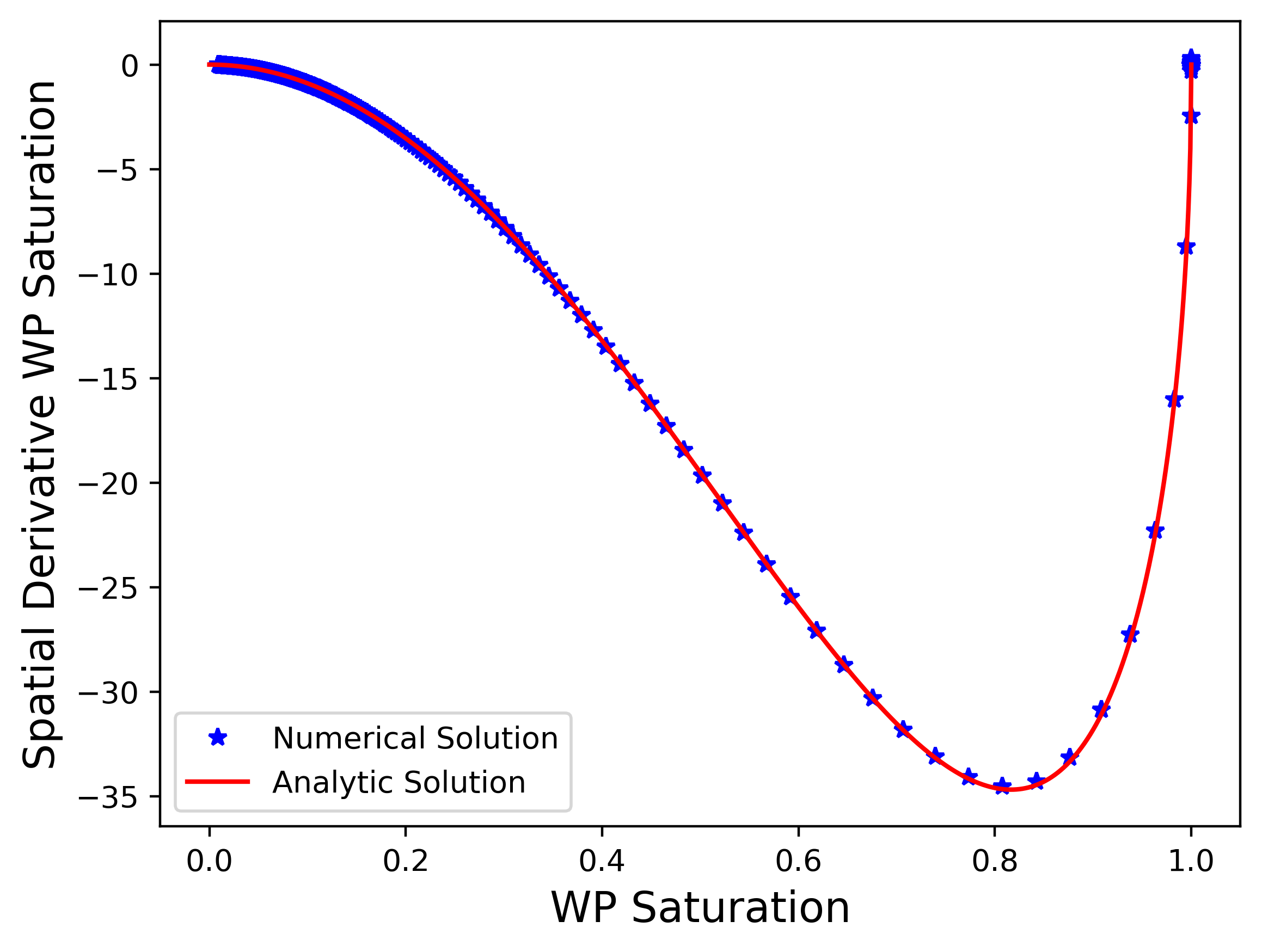}
				\caption{Van Genuchten capillary pressure function}
			\end{subfigure}
			\hfill
			\begin{subfigure}[b]{0.5 \textwidth}
				\centering
				\includegraphics[width=\textwidth]{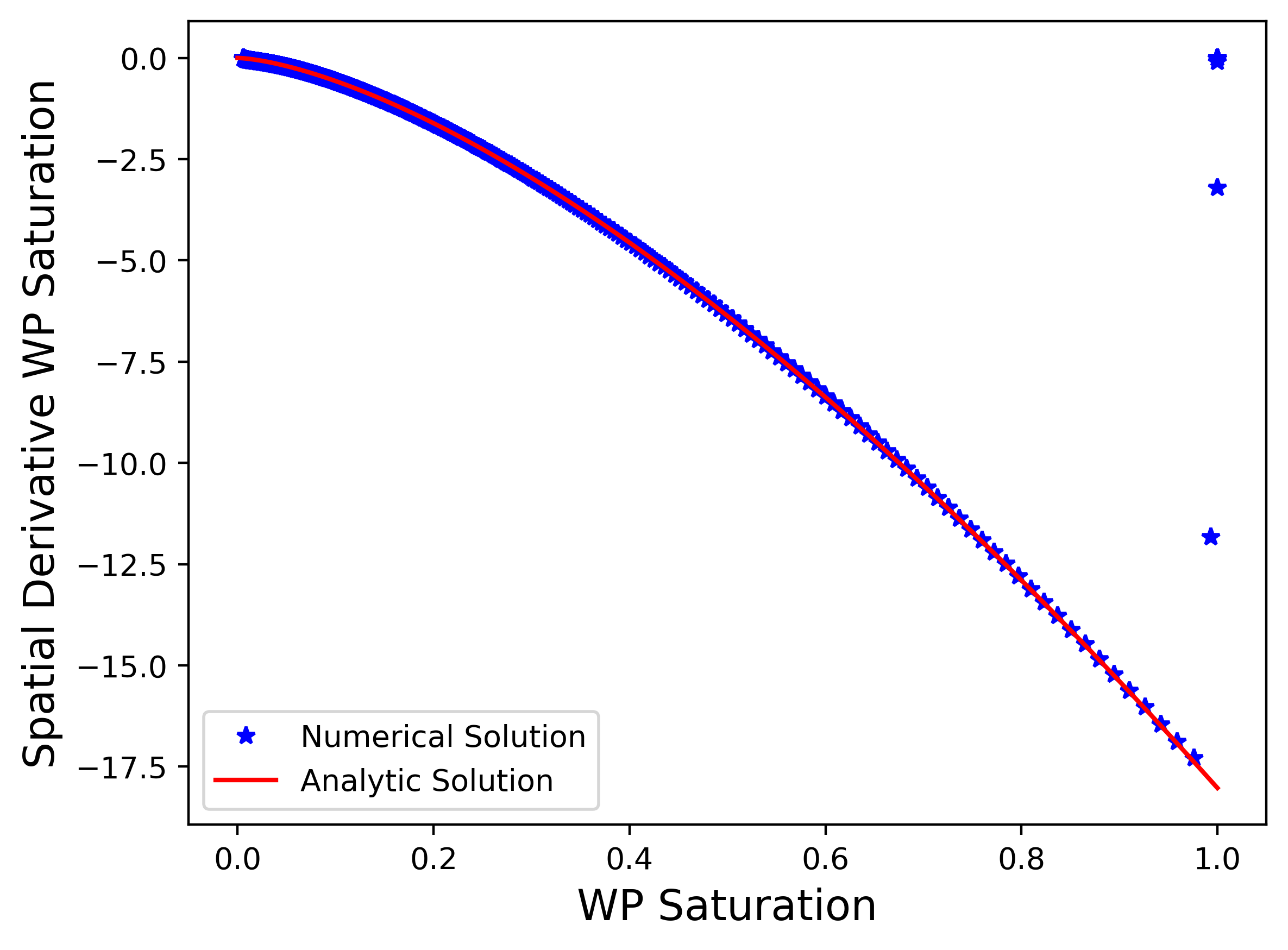}
				\caption{Brooks-Corey capillary pressure function}
			\end{subfigure}
			
			\caption{ These images show the derivatives of the wetting phase saturations with respect to height plotted against the wetting phase saturation. The numerical derivatives are computed from the saturation values obtained in the simulations, as shown in Figure \ref{fig:darcySol_capGravEqSatVsHeight}. The analytic derivatives are calculated by evaluating the right-hand side of Eq. \ref{eq:darcySol_capPressGravEq}. The left image uses the Van Genuchten capillary pressure model, while the right uses the Brooks-Corey model. The grid consists of $10 \times 1000$ voxels. }
			\label{fig:darcySol_capGravEqSatDerivVsHeight}
		\end{figure}
		
		\begin{figure}[ht]
			\begin{subfigure}[b]{0.5\textwidth}
				\centering
				\includegraphics[width=\textwidth]{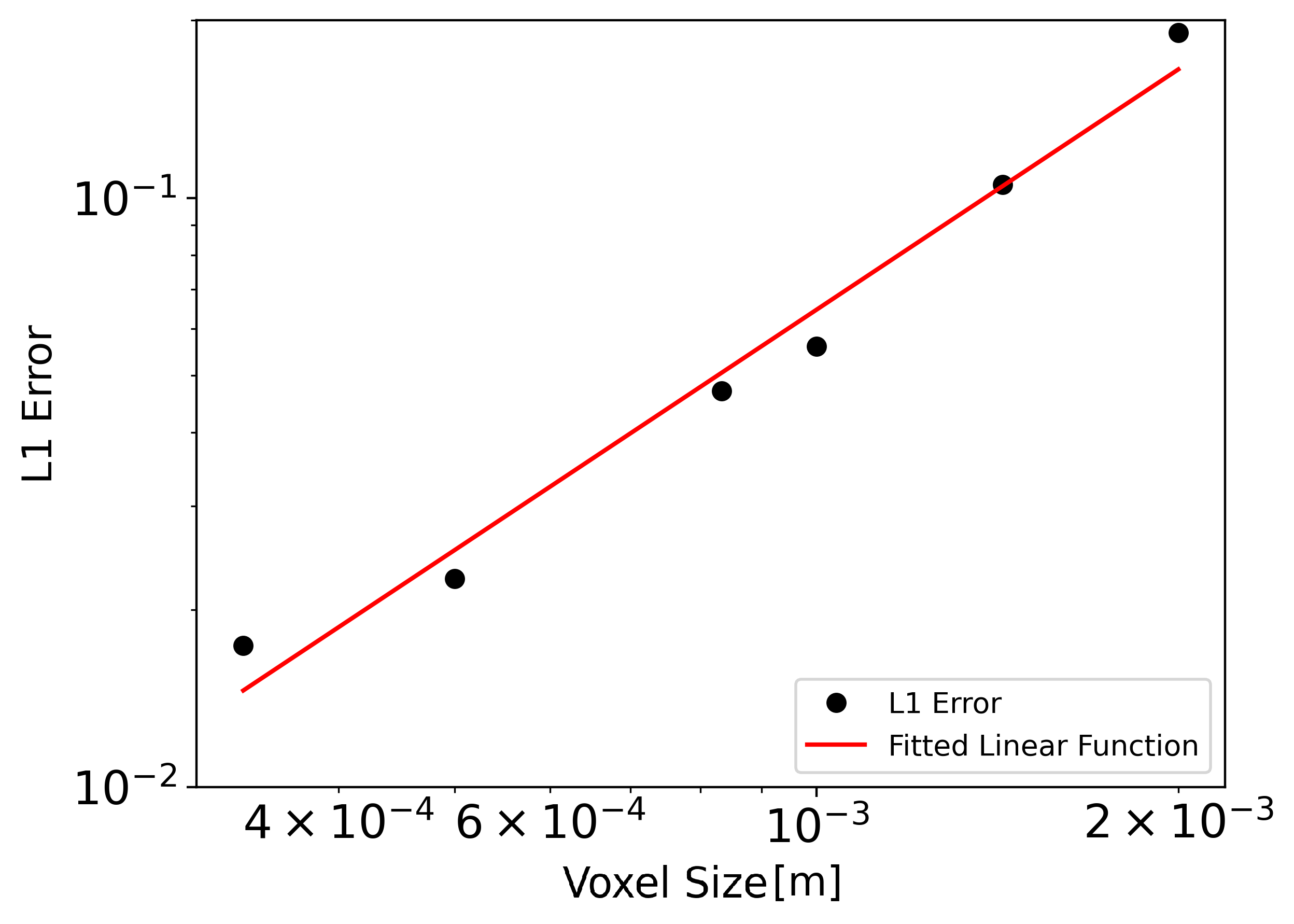}
			\end{subfigure}
			\begin{subfigure}[b]{0.5\textwidth}
				\centering
				\includegraphics[width=\textwidth]{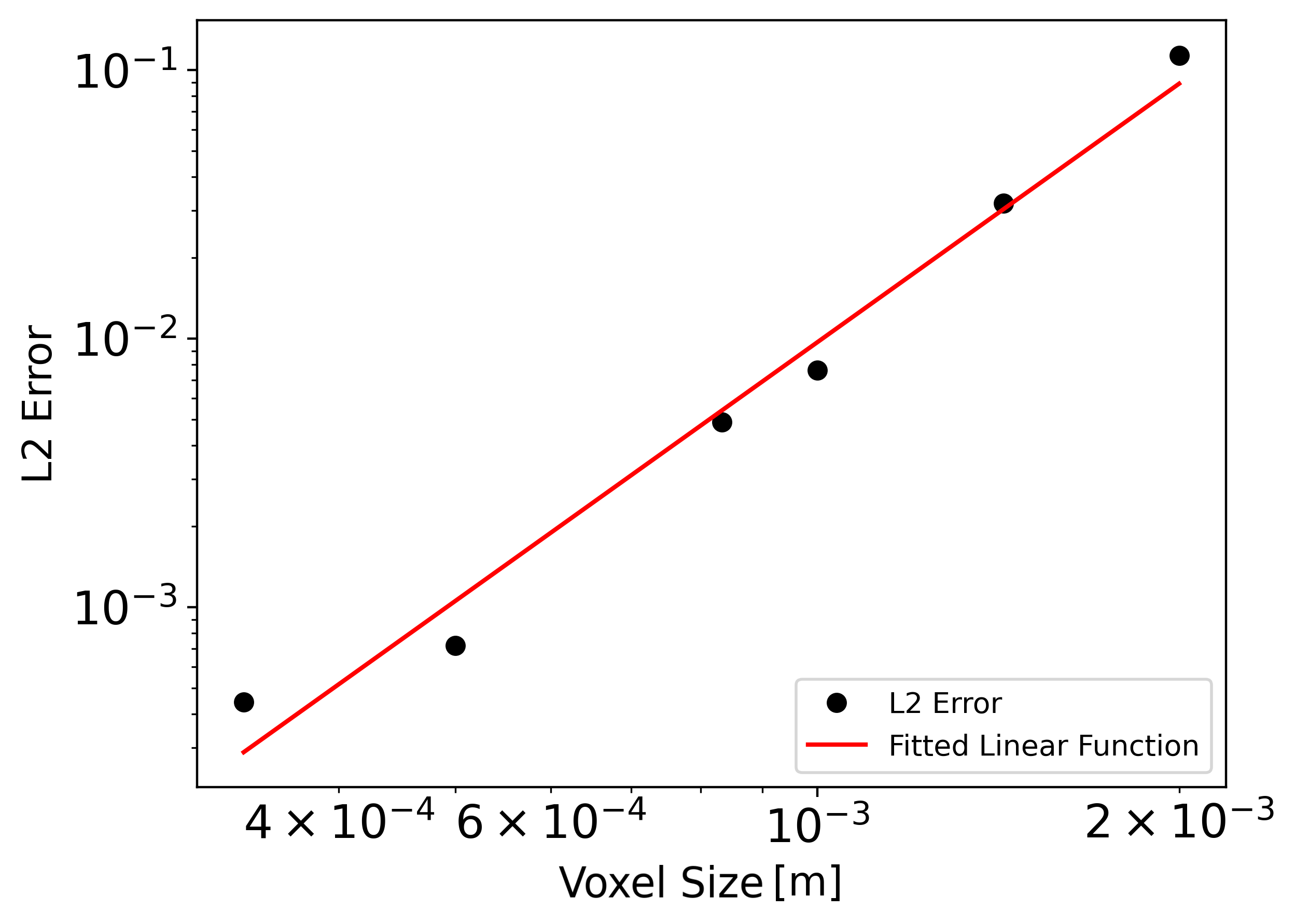}
			\end{subfigure}
			\caption{ These two images illustrate the deviation of the numerically computed derivative of the wetting phase saturation with respect to height from the analytic formula given in Eq. \ref{eq:darcySol_capPressGravEq} for various voxel sizes. The derivatives for a voxel size of $10^{-3} m$ are shown in Figure \ref{fig:darcySol_capGravEqSatDerivVsHeight}. The error is measured with the $L_1$ norm in the left image and the $L_2$ norm in the right image. The Van Genuchten capillary pressure model was used. Numerical errors are represented by black dots, while the red lines correspond to linear least-square regression lines, estimating the order of convergence. The approach to fit these linear regression lines from the error values is the same as in Figure \ref{fig:darcySol_convPlotBL}. The calculated orders of convergence are $1.35$ for the $L_1$ error and $3.20$ for the $L_2$ error. We remark that these are the orders of convergence of the numerically calculated derivative of the saturation with respect to the height and not directly the convergence order of the numerical solution. }
			\label{fig:theory_convPlotCapGravEq}
		\end{figure}
		
		We simulate the setup for $2 \cdot 10^6 s$. Then the system is nearly in its stationary state. \\
		If the system is in its stationary state, it is
		\begin{flalign} \label{eq:darcySol_velStationaryState}
			u_w = u_n = 0 .
		\end{flalign}
		We use Eq. \ref{eq:darcySol_velStationaryState} together with Eq. \ref{two_phase_darcy_comp_2fluid_velEq} for the height direction, that we denote by $h$ in the following formulas. We get
		\begin{flalign} \label{eq:darcySol_capPressGravEqPn}
			\frac{\partial p_n }{ \partial h } &= \rho_n g , \\  \label{eq:darcySol_capPressGravEqPc}
			\frac{\partial p_w }{ \partial h } = \rho_w g \; & \Leftrightarrow \; \frac{\partial p_c }{ \partial h } = \frac{\partial p_n }{ \partial h }  - \rho_w g .
		\end{flalign}
		Inserting Eq. \ref{eq:darcySol_capPressGravEqPn} into Eq. \ref{eq:darcySol_capPressGravEqPc} and using the chain rule leads to
		\begin{flalign} \label{eq:darcySol_capPressGravEq}
			\frac{\partial S_w }{ \partial h } &= \frac{ \left( \rho_n - \rho_w \right) g }{ \frac{\partial p_c }{ \partial S_w } }  .
		\end{flalign}
		By evaluating the right-hand side of Eq. \ref{eq:darcySol_capPressGravEq} at chosen saturation values, the derivative of the saturation field in the height direction can be calculated analytically. The stationary states of the numerical solutions are validated by computing this derivative numerically and comparing it to the analytic result. \\

		\begin{table}[ht]
			\centering
			\caption*{ }
			\begin{tabular}{@{}c|c|c|c|c|c|c|c|c|c@{}}
				$C_{\text{stab}}$ & $ 0.7 $ & $ 0.8 $ & $ 0.9 $ & $ 1.0 $ & $ 1.25 $ & $ 1.5 $ & $ 2 $ & $ 3.0 $ & $ 3.5 $ \\
				\hline
				$L_1$ Error in $[10^{-2}]$ & $ 5.68 $ & $ 5.67 $ & $ 5.66 $ & $ 5.65 $ & $ 5.63 $ & $ 5.60 $ & $ 5.55 $ & $ 5.69 $ & $ 15.0 $    \\
				\hline
				$L_2$ Error in $[10^{-2}]$ & $ 0.77 $ & $ 0.77 $ & $ 0.76 $ & $ 0.76 $ & $ 0.76 $ & $ 0.76 $ & $ 0.75 $  & $ 0.77 $ & $ 4.44 $    \\
				\hline
				Total Iter. Nb. in $[10^{5}]$ & $ 10.86 $ & $ 9.51 $ & $ 8.46 $ & $ 7.61 $ & $ 6.14 $ & $ 5.69 $ & $ 5.13 $ & $ 2.55 $ & $ 2.59 $   \\
				\hline
				Average $ \Delta t$ in $[s]$ & $ 1.84 $ & $ 2.10 $ & $ 2.37 $ & $ 2.63 $ & $ 3.26 $ & $ 3.51 $ & $ 3.90 $ & $ 7.84 $ & $ 7.72 $ 
			\end{tabular}
			\caption*{\textbf{Coats Criterion}}
			
			\begin{tabular}{@{}c|c|c|c|c|c|c|c|c|c@{}}
				$C_{\text{stab}}$ & $ 0.7 $ & $ 0.8 $ & $ 0.9 $ & $ 1.0 $ & $ 1.25 $ & $ 1.5 $ & $ 2 $ & $ 3.0 $ & $ 3.5 $ \\
				\hline
				$L_1$ Error in $[10^{-2}]$ & $ 7.79 $ & $ 9.52 $ & $ 10.5 $ & $ 18.5 $ & $ 23.8 $ & $ 27.4 $ & $ 72.0 $ & $ 22.3 $ & $ 172 $    \\
				\hline
				$L_2$ Error in $[10^{-2}]$ & $ 1.38 $ & $ 1.86 $ & $ 1.86 $ & $ 5.13 $ & $ 8.06 $ & $ 9.57 $ & $ 61.9 $ & $ 8.60 $ & $ 930 $   \\
				\hline
				Total Iter. Nb. in $[10^{5}]$ & $ 2.40 $ & $ 2.81 $ & $ 3.41 $ & $ 3.53 $ & $ 3.37 $ & $ 3.29 $ & $ 3.43 $ & $ 3.75 $ & $ 3.90 $    \\
				\hline
				Average $ \Delta t$ in $[s]$ & $ 8.33 $ & $ 7.11 $ & $ 5.86 $ & $ 5.66 $ & $ 5.93 $ & $ 6.08 $ & $ 5.84 $ & $ 5.33 $ & $ 5.12 $  
			\end{tabular}
			\caption*{\textbf{Characteristic Wave Velocity Criterion}}
			
			\begin{tabular}{@{}c|c|c|c|c|c|c|c|c|c@{}}
				$C_{\text{stab}}$ & $ 0.7 $ & $ 0.8 $ & $ 0.9 $ & $ 1.0 $ & $ 1.25 $ & $ 1.5 $ & $ 2 $ & $ 3.0 $ & $ 3.5 $ \\
				\hline
				$L_1$ Error in $[10^{-2}]$ & $ 5.49 $ & $ 5.81 $ & $ 6.95 $ & $ 6.72 $ & $ 7.94 $ & $ 11.4 $ & $ 24.9 $ & $ 73.3 $ & $ 50.1 $    \\
				\hline
				$L_2$ Error in $[10^{-2}]$ & $ 0.83 $ & $ 1.00 $ & $ 1.33 $ & $ 1.23 $ & $ 1.44 $ & $ 2.10 $ & $ 8.58 $ & $ 65.8 $ & $ 29.6 $    \\
				\hline
				Total Iter. Nb. in $[10^{5}]$ & $ 1.98 $ & $ 2.16 $ & $ 2.31 $ & $ 2.46 $ & $ 2.64 $ & $ 2.70 $ & $ 3.08 $ & $ 3.42 $ & $ 3.36 $   \\
				\hline
				Average $ \Delta t$ in $[s]$ & $ 10.12 $ & $ 9.27 $ & $ 8.66 $ & $ 8.12 $ & $ 7.59 $ & $ 7.40 $ & $ 6.50 $ & $ 5.85 $ & $ 5.95 $  
			\end{tabular}
			\caption*{\textbf{Generalized Characteristic Wave Velocity Criterion}}
						
			\caption{ \label{tab:darcySol_capGravEqTimestepComparison} These tables list the errors in the $L_1$ norm, the errors in the $L_2$ norm, the total number of time iterations, and the arithmetic mean of all timestep sizes for simulations of the capillary gravity equalization problem. The $L_1$ and $L_2$ errors are determined by the subtraction of the analytic and numerical derivatives of the wetting phase saturation with respect to height in the stationary state. The first table uses the Coats criterion, the second table uses the characteristic wave velocity criterion, and the last table uses the generalized characteristic wave velocity criterion. In each case, a range of values for the stability constant $C_{\text{stab}}$ is tested. All simulations were performed with $\tau_{\text{max}} = 0.3$, an initial timestep size of $1 \cdot 10^{-3} s$ and a grid consisting of $10 \times 1000$ voxels. } 
		\end{table}

		Figure \ref{fig:darcySol_capGravEqSatVsHeight} displays the stationary saturation fields of the numerical solutions for both cases and in Figure \ref{fig:darcySol_capGravEqSatDerivVsHeight} the saturation derivatives with respect to height, calculated numerically and analytically, are presented. This comparison illustrates a strong agreement between numerical and analytic stationary states in both cases. \\
		In the Brooks-Corey scenario, the numerical derivative differs from the analytic one for saturation values close to one. This difference arises from the nearly constant saturation region near height zero and the sharp edge at its end in the numerical solution. Within this region, the saturation shows a small gradient leading to the sharp edge, which causes the small derivative values near saturation one. This effect is not present in the analytic solution. These deviations might decrease for extended simulation times, as the numerical solutions slowly approach their stationary state. \\
		
		Figure \ref{fig:theory_convPlotCapGravEq} illustrates the deviation of the numerical saturation derivative from the analytic derivative for various voxel sizes, using the Van Genuchten capillary pressure model. We estimate convergence orders of $1.35$ for the $L_1$ error and $3.20$ for the $L_2$ error. Note that these convergence orders refer to the saturation derivative with respect to height rather than the numerical solution itself. \\

		We similarly compare the timestep criteria as in the Buckley-Leverett section. The error and timestep statistics of these simulations are provided in Table \ref{tab:darcySol_capGravEqTimestepComparison} to evaluate the accuracy and computational cost. Unlike the Buckley-Leverett example, here the $L_1$ and $L_2$ errors are computed based on the difference between the numerical and analytic derivatives of the wetting phase saturation with respect to height in the stationary state. \\
		The generalized characteristic wave velocity criterion has notably lower errors and fewer time iterations compared to the characteristic wave velocity criterion. \\
		When comparing the Coats criterion to the generalized characteristic wave velocity criterion, the Coats criterion yields slightly smaller errors but requires substantially more time iterations. For instance, at $C_{\text{stab}} = 1$, the $L_1$ error of the Coats criterion is approximately $16 \%$ lower but demands over $3$ times the number of time iterations. Lowering $C_{\text{stab}}$ to $0.7$ or $0.8$ produces similar errors for both criteria. Increasing the stability constant of the Coats criterion, for example to $C_{\text{stab}} = 3$, can bring its time iteration count close to that of the generalized characteristic wave velocity criterion at $C_{\text{stab}} = 1$. But again this requires prior knowledge of the appropriate stability constant, which is rarely available in practical applications. \\
		
		In conclusion, the solver accurately reproduces the known stationary state of the flow that is driven by gravity and capillary forces. Furthermore, the generalized characteristic velocity criterion effectively limits the timestep size in this example, requiring significantly fewer time iterations than the other timestep criteria, while having a good agreement with the analytic stationary state.  \\
		Comparing runtimes again, the generalized characteristic wave velocity criterion with $C_{\text{stab}} = 1$ takes $12,284s$, while the Coats criterion takes $32,231s$. Thus, the runtime of the Coats criterion is approximately $2.6$ times longer than that of the generalized characteristic wave velocity criterion. This is a slightly less than the three times as many time iterations, but still represents a large improvement.

		\subsection{Compression of Gas} \label{sec:darcySol_compressGas}
		
\begin{figure}[ht]
			\centering
			\includegraphics[width=0.6\textwidth]{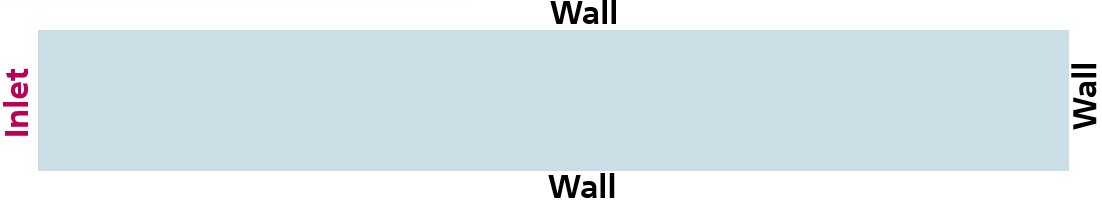}
			\caption{This two-dimensional geometry measures $1 \times 0.1 m$. The boundary on the left is an inlet. All other boundaries are walls, meaning no flow can pass through them.}
			\label{fig:darcySol_compressGasGeo}
		\end{figure}

		In this section, we use the two-dimensional geometry shown in Figure \ref{fig:darcySol_compressGasGeo}, measuring $1m$ in width and $0.1m$ in length. The left boundary is an inlet boundary and the other three boundaries are wall boundaries. One boundary serves as an inlet, while the other three are walls. Initially, the geometry is filled with air at atmospheric pressure, i.e., $10^5 Pa$. Capillary pressure and gravity are neglected in this example. At the inlet, various pressure values $p_I$ are prescribed as Dirichlet boundary conditions, all greater than $10^5 Pa$. As a result, the wetting phase enters the porous medium by compressing the air inside. Without capillary pressure and gravity effects, the wetting phase does not spread through the porous medium. Consequently, the gas compression continues until its internal pressure matches the inlet pressure. \\
		
		\begin{table}[ht]
			\centering
			\begin{tabular}{@{}c|c|c|c|c|c|c@{}}
				$\phi$ & $K_0$  &  $\mu_w$ &  $\mu_n$ &  $\rho^{ref}_n$ &  $p^{ref}_n$ & $R_n$  \\
				\hline
				$0.6$ & $4 \cdot 10^{-12} m^2$ & $1 \cdot 10^{-4} Pa s $ & $1.76 \cdot 10^{-5} Pa s $ & $ 1.22 \frac{kg}{m^3}$ & $10^{5} Pa$ & $10^5 Pa$   \\
			\end{tabular}
			\caption{ The ideal gas law, Eq. \ref{idealGasLaw}, is used to model the compressibility of the non-wetting phase. The wetting phase is assumed to be incompressible.  } 
			\label{tab:darcySol_paramsCompressGas}
		\end{table}
		
		In this example, the physical parameters listed in Table \ref{tab:darcySol_paramsCompressGas} are used, and the ideal gas law, Eq. \ref{idealGasLaw}, is applied to model the compressibility of air. Simplified Brooks-Corey relative permeabilities with $m^{w,BC} = m^{nw,BC} = 4$ are employed and all simulations in this section utilize a grid of the size $1000 \times 10$ voxels. \\
		The area occupied by the wetting phase in the stationary state can be determined by calculating the compression of the non-wetting phase at the inlet pressure $p_I$. For this, we divide the non-wetting phase density at the initial pressure by the non-wetting phase density at the inlet pressure $p_I$, resulting in 
		\begin{flalign} \label{eq:darcySol_compressGasDensityFrac}
			\frac{\rho_n(10^5 Pa)}{\rho_n(p_I)} \overset{ \text{Eq. } \ref{idealGasLaw}}{=} \frac{\rho_n^{ref}}{\rho_n^{ref} + \frac{1}{R_n} \left( p_I - 10^5 Pa \right) \rho_n^{ref}} = \frac{1}{1 + \frac{p_I - 10^5 Pa }{R_n}} \overset{\text{Table \ref{tab:darcySol_paramsCompressGas}}}{=} \frac{10^5 Pa}{p_I}.
		\end{flalign}
		For example, when $p_I = 2 \cdot 10^5 Pa$, the ratio from Eq. \ref{eq:darcySol_compressGasDensityFrac} equals $0.5$, meaning the non-wetting phase compresses to half of its original area, allowing the wetting phase to fill the remaining half of the geometry. \\

		\begin{figure}[ht]
			\centering
			\includegraphics[width=0.6\textwidth]{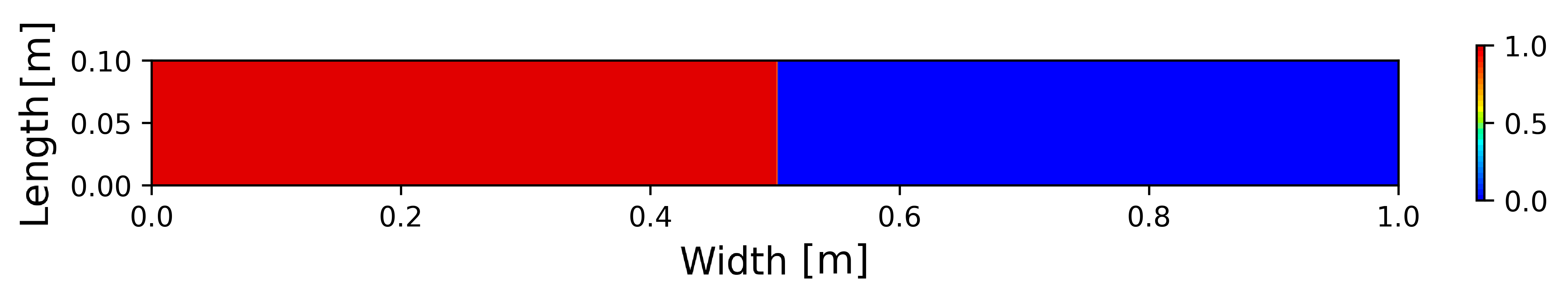}
			\caption{ This image displays the wetting phase saturation field from the numerical solution with an inlet pressure of $2 \cdot 10^5 Pa$ after $2500 s$. At this time, the solution has reached the stationary state. The simulation uses $m_I = 5$.  }
			\label{fig:darcySol_compressGasResult2Bar}
		\end{figure}
		
		\begin{figure}[ht]
			\centering
			\includegraphics[width=0.65\textwidth]{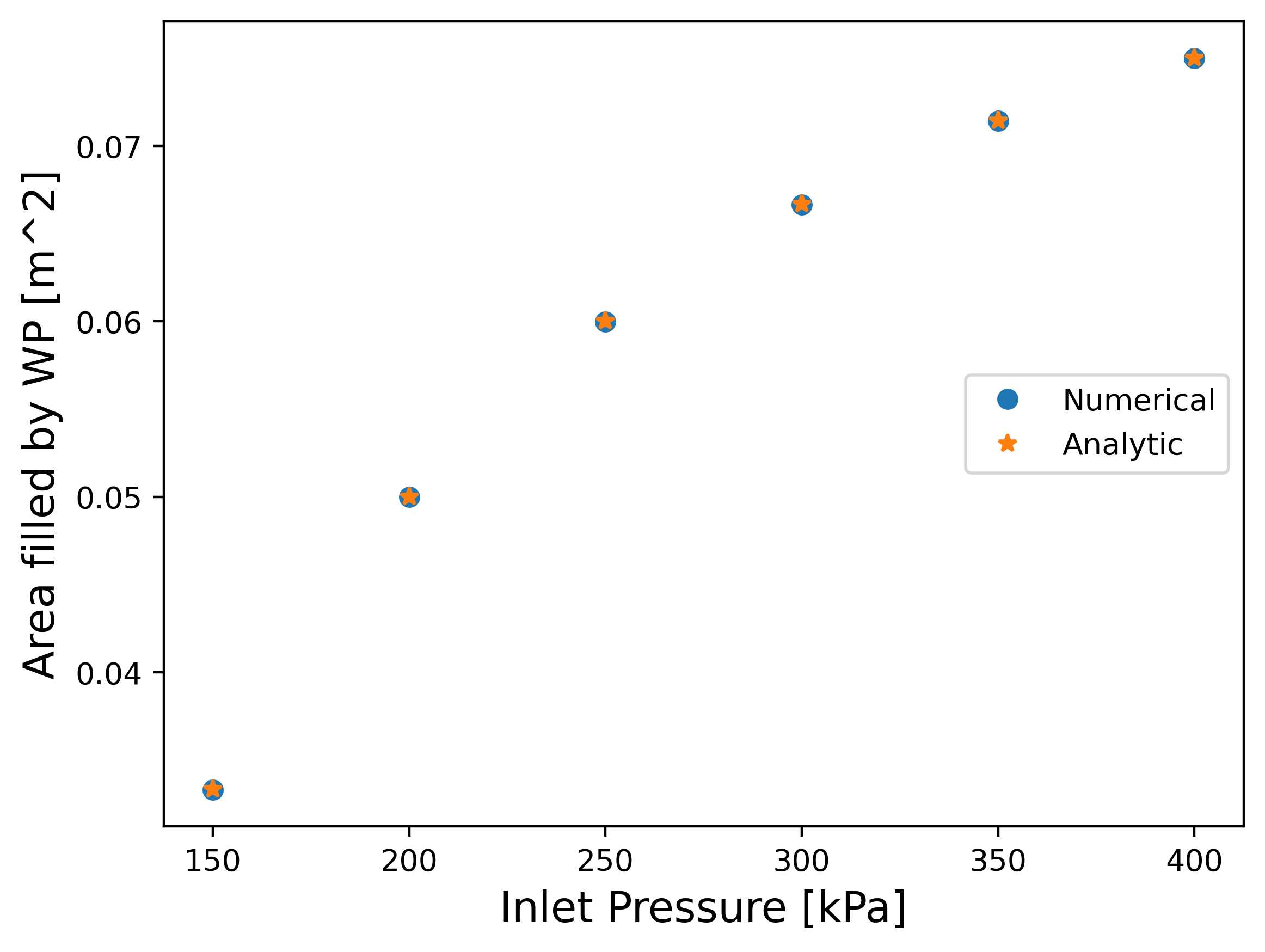}
			\caption{ This image compares the numerically and analytically determined areas occupied by the wetting phase in the stationary state for various inlet pressures $p_I$. The simulations are performed with $m_I = 5$. }
			\label{fig:darcySol_compressGasInletPressVsSat}
		\end{figure}
		
		Figure \ref{fig:darcySol_compressGasResult2Bar} shows the saturation field of the numerical solution in the stationary state with an inlet pressure of $2 \cdot 10^5 Pa$. According to Eq. \ref{eq:darcySol_compressGasDensityFrac}, the wetting phase is expected to occupy half of the geometry, which closely matches the numerical result. \\
		
		In Figure \ref{fig:darcySol_compressGasInletPressVsSat}, the stationary wetting phase areas from numerical solutions with various inlet pressures $p_I$ are compared to the analytic predictions, showing a good agreement between them. \\
		
		\begin{table}[ht]
			\centering
			\begin{tabular}{@{}c|c|c|c|c|c|c|c@{}}
				&  $m_I = 1$  & $m_I = 2$  & $m_I = 3$ & $m_I = 4$  & $m_I = 5$ & $m_I = 6$  & $m_I = 7$  \\
				\hline
				\textbf{Total} &   & & & & & &  \\
				\textbf{Mob.} & $ 1.36 \cdot 10^{-3}$ & $ 1.04 \cdot 10^{-2} $ & $ 1.22 \cdot 10^{-4}$ & $1.26 \cdot 10^{-3} $ &   $ 3.43 \cdot 10^{-5}$ & $4.62 \cdot 10^{-4} $ & $7.65 \cdot 10^{-5} $ \\
				\hline
				\textbf{Phase} &   & & & & & &  \\
				\textbf{Mob.} & $ 1.57 \cdot 10^{-3}$ & $ 4.00 \cdot 10^{-2} $ & $ 8.43 \cdot 10^{-4}$ & $ 1.30 \cdot 10^{-3} $ &   $ 5.24 \cdot 10^{-4}$ & $ 1.43 \cdot 10^{-3} $ & $ 3.48 \cdot 10^{-4} $ \\
			\end{tabular}
			\caption{ This table shows the relative errors in the non-wetting phase mass for different numbers of IMPES iterations $m_I$ and two discretizations. The "Total Mobility" approach employs $\overline{M K_0}^{H,f_{ij}}$ in the discrete total velocity $u$, whereas the "Phase Mobility" approach uses $\overline{M_w K_0}^{H,f_{ij}} + \overline{M_n K_0}^{H,f_{ij}}$. Both discretizations were discussed in Section \ref{sec_pressEq}. } 
			\label{tab:darcySol_compressGasNWPMassVsIMPESIter}
		\end{table}	
		
		The non-wetting phase mass should remain constant over time since it is trapped within the geometry. However, numerical errors cause variations in the non-wetting phase mass. Table \ref{tab:darcySol_compressGasNWPMassVsIMPESIter} lists the relative deviations of the non-wetting phase mass from its initial value for $m_I = 1$ to $m_I = 7$, where $m_I$ is the number of IMPES iterations. The table also compares the non-wetting phase mass errors for two discretizations of the total velocity $u$. Both discretizations are explained in Section \ref{sec_pressEq}. \\
		In this section, relative deviations are computed by evaluating the non-wetting phase mass at $101$ equally distributed timesteps between $0$ and $2500 s$. This non-wetting phase mass is then subtracted by the initial non-wetting phase mass, and divided by the initial non-wetting phase mass. The maximum deviation among these times is reported as the overall deviation of the simulation. \\
		Table \ref{tab:darcySol_compressGasNWPMassVsIMPESIter} shows a superior mass conservation of the "Total Mobility" discretization, which is why we generally use this discretization in the IMPES solver. \\
		Furthermore, Table \ref{tab:darcySol_compressGasNWPMassVsIMPESIter} reveals that increasing the number of IMPES iterations generally decreases the error in non-wetting phase mass conservation up to $m_I = 5$. Beyond this, the error does not decrease further or may even increase. The iterative method also performs better for odd values of $m_I$. The errors for even values still decrease with increasing $m_I$, but the errors are larger than those of the preceding odd value of $m_I$. We do not have a definitive explanation for this behavior. It is possible that the solution oscillates around the correct total mass value with each IMPES iteration, and that the overshoots at even values of $m_I$ are larger than those at odd values. \\
		
		\begin{table}[ht]
			\centering
			\caption*{}
			\begin{tabular}{@{}c|c|c|c|c|c|c|c|c@{}}
				$C_{\text{stab}}$ & $ 0.8 $ & $ 0.9 $ & $ 1 $ & $ 1.1 $ & $ 1.2 $ & $ 1.3 $ & $ 1.4 $ & $ 1.5 $ \\
				\hline
				Rel. Error NWP Mass in $[10^{-4}]$ & $ 1.18 $ & $ 3.38 $ & $ 32.9 $ & $ 362 $ & $ 1247 $ & $ 542 $ & $ 475 $  & $ 708 $   \\
				\hline
				Total Iter. Nb. & $ 5850 $ & $ 5354 $ & $ 4334 $ & $ 3333 $ & $ 3795 $ & $ 3182 $ & $ 3405 $ & $ 3317 $   \\
				\hline
				Average $ \Delta t$ in $[s]$ & $ 0.43 $ & $ 0.47 $ & $ 0.58 $ & $ 0.75 $ & $ 0.66 $ & $ 0.79 $ & $ 0.73 $ & $ 0.75 $
			\end{tabular}
			\caption*{\textbf{Coats Criterion}}
			
			\begin{tabular}{@{}c|c|c|c|c|c|c|c|c@{}}
				$C_{\text{stab}}$ & $ 0.8 $ & $ 0.9 $ & $ 1 $ & $ 1.1 $ & $ 1.2 $ & $ 1.3 $ & $ 1.4 $ & $ 1.5 $   \\
				\hline
				Rel. Error NWP Mass in $[10^{-4}]$ & $ 0.99 $ & $ 0.75 $ & $ 1.90 $ & $ 3.27 $ & $ 4.72 $ & $ 5.99 $ & $ 4.32 $ & $ 37.1 $    \\
				\hline
				Total Iter. Nb. & $ 5419 $ & $ 5045 $ & $ 4748 $ & $ 4506 $ & $ 4306 $ & $ 4138 $ & $ 3997 $ & $ 3879 $   \\
				\hline
				Average $ \Delta t$ in $[s]$ & $ 0.46 $ & $ 0.50 $ & $ 0.53 $ & $ 0.55 $ & $ 0.58 $ & $ 0.60 $ & $ 0.63 $ & $ 0.64 $
			\end{tabular}
			\caption*{\textbf{Characteristic Wave Velocity Criterion}}
			
			\begin{tabular}{@{}c|c|c|c|c|c|c|c|c@{}}
				$C_{\text{stab}}$ & $ 0.8 $ & $ 0.9 $ & $ 1 $ & $ 1.1 $ & $ 1.2 $ & $ 1.3 $ & $ 1.4 $ & $ 1.5 $ \\
				\hline
				Rel. Error NWP Mass in $[10^{-4}]$ & $ 1.02 $ & $ 0.71 $ & $ 1.86 $ & $ 3.20 $ & $ 4.71 $ & $ 5.86 $ & $ 4.42 $ & $ 36.6 $    \\
				\hline
				Total Iter. Nb. & $ 5420 $ & $ 5046 $ & $ 4749 $ & $ 4507 $ & $ 4307 $ & $ 4140 $ & $ 3997 $ & $ 3879 $   \\
				\hline
				Average $ \Delta t$ in $[s]$ & $ 0.46 $ & $ 0.50 $ & $ 0.53 $ & $ 0.55 $ & $ 0.58 $ & $ 0.60 $ & $ 0.63 $ & $ 0.64 $
			\end{tabular}
			\caption*{\textbf{Generalized Characteristic Wave Velocity Criterion}}
			
			\caption{ \label{tab:darcySol_compressGasTimestepComparison} These tables list the relative error in the mass of then non-wetting phase, the total number of time iterations, and the arithmetic mean of all timestep sizes for simulations of the compression of gas example. The first table uses the Coats criterion, the second table uses the characteristic wave velocity criterion, and the last table uses the generalized characteristic wave velocity criterion. In each case, a range of values for the stability constant $C_{\text{stab}}$ is tested. All simulations were performed with $\tau_{\text{max}} = 0.3$, an initial timestep size of $1 \cdot 10^{-4} s$, and a grid consisting of $1000 \times 10$ voxels. }
		\end{table}
		
		Table \ref{tab:darcySol_compressGasTimestepComparison} compares the three timestep criteria. It presents the total number of time iterations, average timestep size, and, for this example, the relative deviation of the non-wetting phase mass for various values of the stability constant $C_{\text{stab}}$. \\
		For this example the characteristic wave velocity criterion and the generalized characteristic wave velocity criterion again lead to similar results. At $C_{\text{stab}} = 1$, the Coats criterion results in a relative error in non-wetting phase mass about $17.7$ times larger than that of the generalized characteristic wave velocity criterion, while requiring only about $8.8 \%$ fewer time iterations. Although the Coats criterion can achieve similarly low errors at smaller stability constants, the total number of iterations then exceeds that of the generalized characteristic wave velocity criterion at $C_{\text{stab}} = 1$. \\		
		This demonstrates that the solver and the newly proposed timestep criterion are applicable to compressible fluid phases. For compressible phases, using $m_I = 5$ is recommended to reduce errors in the non-wetting phase mass. The new timestep criterion using $C_{\text{stab}} = 1$ also results in significantly lower non-wetting phase mass errors compared to the Coats criterion.

		\subsection{Discontinuity of Material Parameters} \label{sec:darcySol_discontMat}
		
\begin{figure}[ht]
			\centering
			\includegraphics[width=0.65\textwidth]{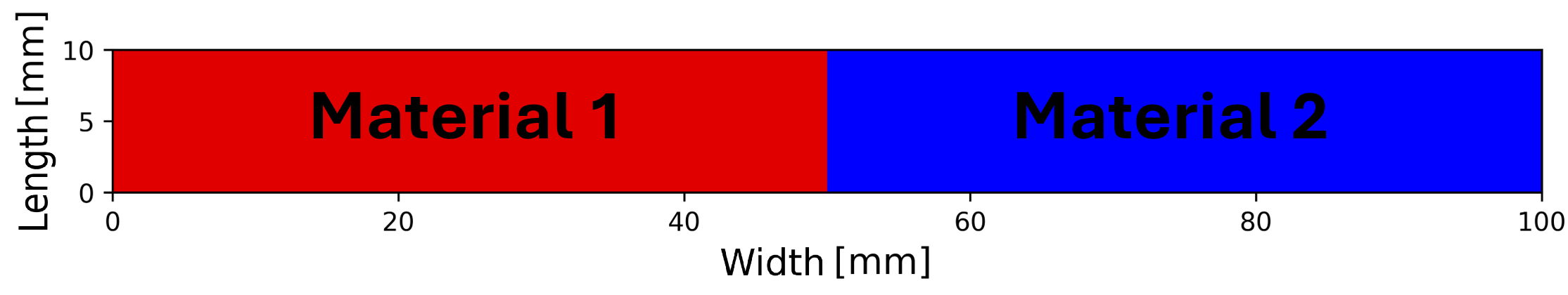}
			\caption{ The geometry consists of two adjacent porous media, resulting in discontinuities in two-phase flow material parameters across the interface. }
			\label{fig:darcySol_discontMatGeo}
		\end{figure}

		In this section, the solver is tested on an example featuring discontinuous material parameters. \\
		The geometry depicted in Figure \ref{fig:darcySol_discontMatGeo} is used, with different material parameters assigned to the two porous media. Wall boundary conditions are applied on all outer boundaries, preventing any fluid from entering or leaving the geometry. The gravity is neglected, and the fluid phases are assumed to be incompressible. \\
		
		\begin{table}[ht]
			\centering
			\begin{tabular}{@{}c|c|c|c|c|c@{}}
				$\phi_1$  & $\phi_2$ & $K_{0,1}$  & $K_{0,2}$ &  $\mu_w$ &  $\mu_n$  \\
				\hline
				$0.42$ & $0.5$ &  $5.5 \cdot 10^{-11} m^2 $ & $5.8 \cdot 10^{-11} m^2 $ & $ 6.72 \cdot 10^{-2} Pa s $ & $ 1.76 \cdot 10^{-5} Pa$   \\
			\end{tabular}
			\caption{ This table summarizes the fluid and porous media parameters used in the experiments of Section \ref{sec:darcySol_discontMat}. The subscripts $1$ and $2$ denote material parameters corresponding to material $1$ and material $2$, respectively. Both fluid phases are assumed to be incompressible. } 
			\label{tab:darcySol_paramsDiscontMat}
		\end{table}
		
		\begin{figure}[ht]
			\centering
			\includegraphics[width=0.5\textwidth]{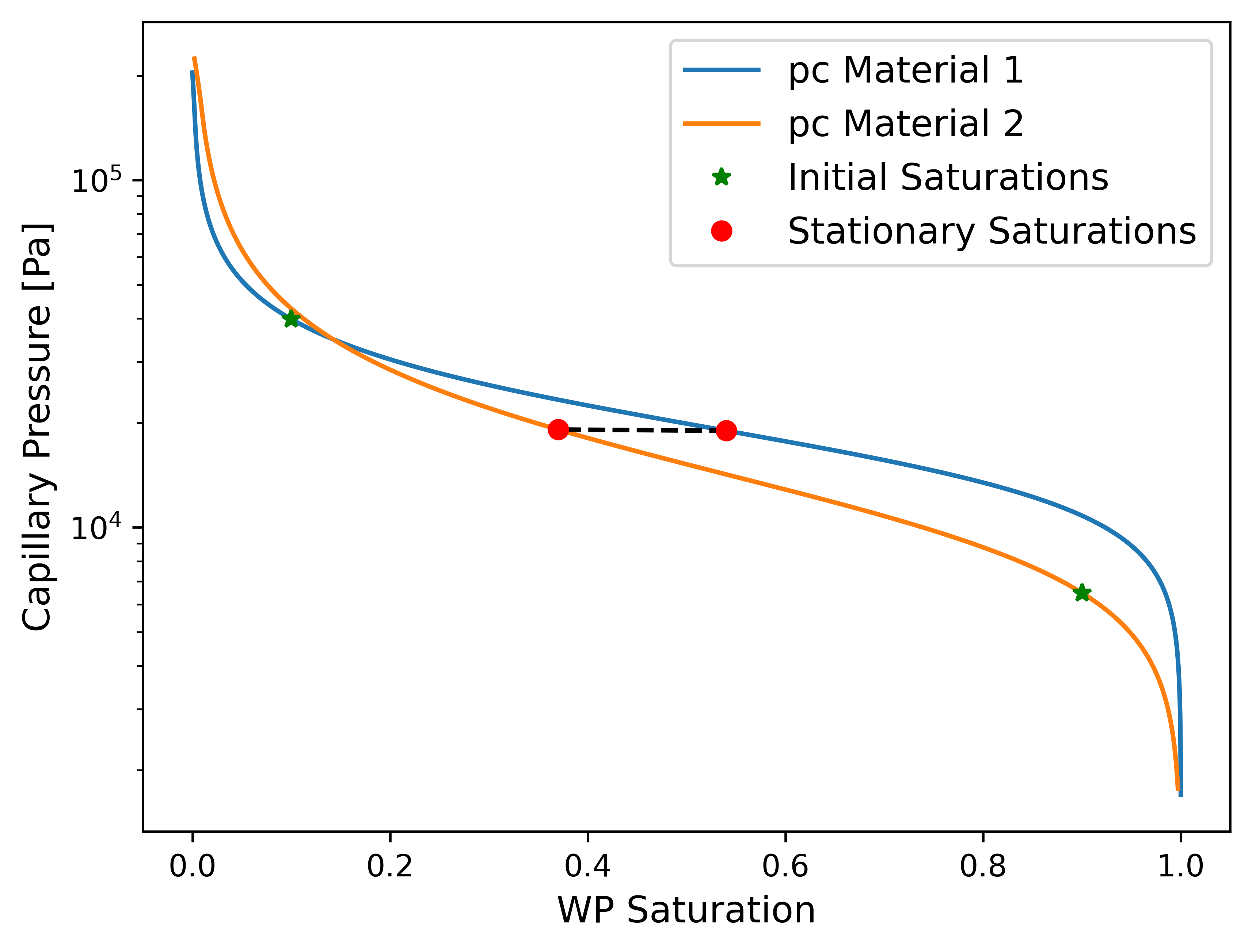}
			\caption{ This image depicts the capillary pressure functions for the porous media in the geometry shown in Figure \ref{fig:darcySol_discontMatGeo}. The solid blue line corresponds to the capillary pressure function of material $1$, while the solid orange line represents that of material $2$. Green stars indicate the initial saturation states for both materials, and red circles mark the stationary states. In the stationary state, the capillary pressures in both porous materials are equal. }
			\label{fig:darcySol_discontMatCapPressFcts}
		\end{figure}
		
		The parameters listed in Table \ref{tab:darcySol_paramsDiscontMat} are used. The non-wetting phase is air and the wetting phase correspond to a light polymer resin. For porous material $1$, the simplified Brooks-Corey relative permeability functions with $m_1^{w,BC} = m_1^{nw,BC} = 4$ and a Van Genuchten capillary pressure function with $p_{e,1}^{VG} = 17.7 kPa$ and $m_1^{VG} = 0.74$ are used. For porous material $2$, the simplified Brooks-Corey relative permeability functions with $m_2^{w,BC} = m_2^{nw,BC} = 3$ and a Van Genuchten capillary pressure function with $p_{e,2}^{VG} = 12 kPa$ and $m_2^{VG} = 0.64$ are used. Initial saturations are set to $0.1$ in material $1$ and $0.9$ in material $2$. Both capillary pressure functions and initial states are shown in Figure \ref{fig:darcySol_discontMatCapPressFcts}. \\

		\begin{figure}[ht]
			\begin{subfigure}[b]{0.5\textwidth}
				\centering
				\includegraphics[width=\textwidth]{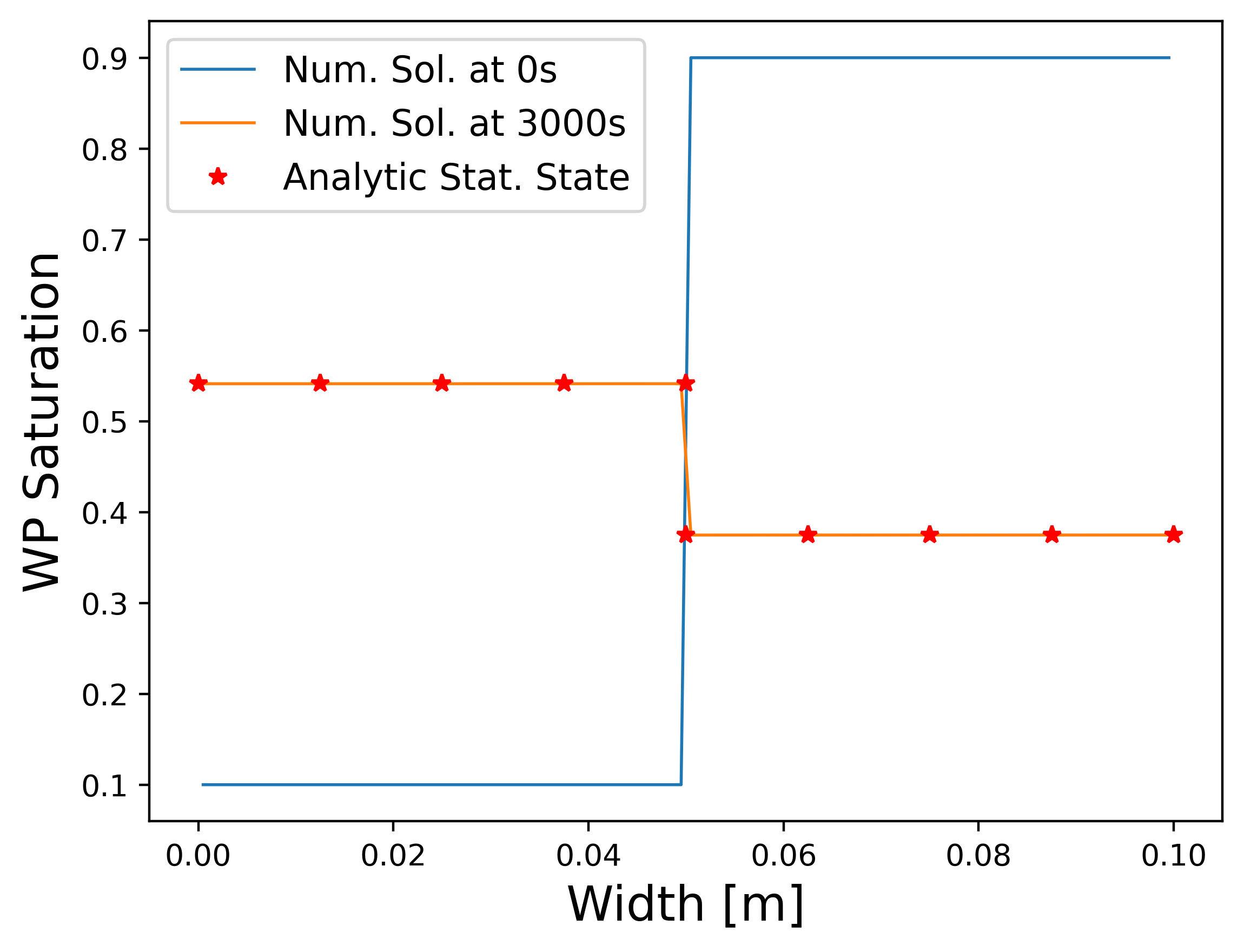}
			\end{subfigure}
			\hfill
			\begin{subfigure}[b]{0.5 \textwidth}
				\centering
				\includegraphics[width=\textwidth]{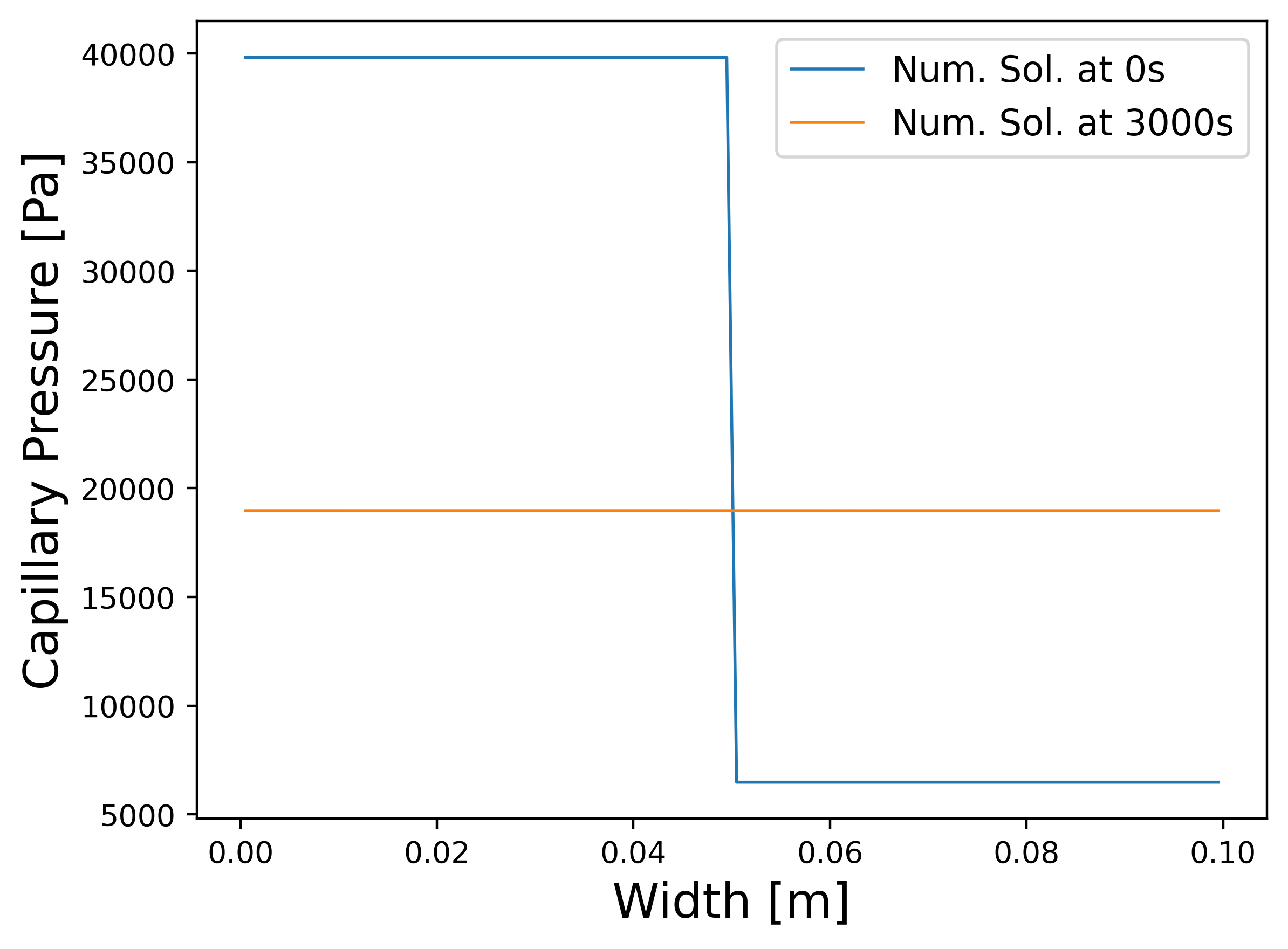}
			\end{subfigure}		
			
			\caption{ The right image displays the saturation profile along the width direction. The blue line corresponds to the initial state, while the orange line shows the state after $3000s$ of simulation. The analytically calculated saturations in the stationary state are marked by red stars. The left image presents the capillary pressures at the same times. The simulation uses a grid of $1000 \times 10$ voxels.  }
			\label{fig:darcySol_discontMatResults}
		\end{figure}
		
		With this setup, the capillary pressure is initially discontinuous at the interface. Since there are no external influences on the flow, the capillary pressure becomes uniform across the interface in the stationary state. Because the two materials have different capillary pressure functions, equal capillary pressure at the interface results in a saturation jump. The stationary saturations $\hat{S}_{w,1}, \hat{S}_{w,2} \in [0, 1]$ in materials $1$ and $2$ are characterized by
		\begin{flalign} \label{eq:darcySol_discontMatCapPressEq}
			p_{c,1}(\hat{S}_{w,1}) &= p_{c,2}(\hat{S}_{w,2}) , \\  \label{eq:darcySol_discontMatMassConserv}
			\phi_1 \hat{S}_{w,1} + \phi_2 \hat{S}_{w,2} &= \phi_1 \cdot 0.1 + \phi_2 \cdot 0.9 .
		\end{flalign}
		Eq. \ref{eq:darcySol_discontMatCapPressEq} ensures the stationary state is reached as the capillary pressure is homogeneous across the interface, while Eq. \ref{eq:darcySol_discontMatMassConserv} enforces the mass conservation. It requires the phase saturations to remain constant due to the wall boundary conditions and the incompressibility of the fluid phases. \\
		Solving the system of Eq. \ref{eq:darcySol_discontMatCapPressEq} and Eq. \ref{eq:darcySol_discontMatMassConserv} yields stationary saturations of approximately $\hat{S}_{w,1} \approx 0.54$ and $\hat{S}_{w,2} \approx 0.37$. These values are shown in Figure \ref{fig:darcySol_discontMatCapPressFcts} together with the capillary pressure functions of both porous materials. \\
		Figure \ref{fig:darcySol_discontMatResults} displays the saturation and capillary pressure fields from the simulation after $3000s$, showing strong agreement with the analytically determined stationary state. \\
		This example illustrates that the solver effectively handles discontinuous material parameters.

	\section{Summary and Conclusions}
	
		A numerical method to simulate the two-phase Darcy equations was introduced. The method is an IMPES method with a finite volume discretization and a newly presented timestep criterion. \\
		
		The solver can reliably simulate pressure drop and capillary dominated flows. This is validated using the Buckley-Leverett problem and an equalization problem of capillary pressure and gravity forces with a known stationary state. Moreover, the solver can simulate with compressible fluid phases and with discontinuous material parameters. Both is tested on examples with known stationary states. As the solver decouples the pressure and saturation equation the mass conservation of the non-wetting phase is critical. We showed that the error in the conservation of the non-wetting phase mass is low and it can be further reduced by using multiple IMPES iterations in every timestep. The results indicate that increasing the number of IMPES iterations is effective up to $5$ and that odd numbers yield better accuracy than even numbers.  \\
		
		The new timestep criterion extends the one used by \cite{lamine2015multidimensional} to include flows with capillary effects. The proposed timestep criterion uses numerical approximations of the derivatives of $u$ and $u_D$ with respect to the saturation to appropriately restrict the timestep in all flow scenarios. \\
		With this timestep criterion all presented examples were efficiently and accurately simulated using $C_{\text{stab}} = 1$, independent what the dominant flow effects are. The comparisons with the Coats criterion showed that the presented timestep criterion achieves similar accuracy with fewer time iterations. Specifically, for the Buckley-Leverett example, the Coats criterion required about $19 \%$ more iterations, while for the capillary pressure gravity equalization example, it needed over three times as many. In both examples, comparison of runtimes on the same hardware lead to similar reductions as in the comparison of needed time iterations. This demonstrates that the generalized characteristic wave velocity criterion is not significantly more complex to compute than the Coats criterion. \\
		In the case of compressible fluid phases, the new criterion produced significantly smaller errors in the non-wetting phase mass conservation compared to the Coats criterion. Overall, the proposed timestep criterion reliably limits the timestep sizes for the IMPES solver across various flow scenarios and proves to be more efficient than the Coats criterion in the tested examples.
		
		\section*{Notation}
		
		\begin{longtable}{@{}lr@{}}
			$\phi$  & porosity   \\
			$ \rho_d $  & density of phase $d \in \{ w, n \}$   \\
			$ S_d $  & saturation of phase $d \in \{ w, n \}$   \\
			$ u_d $  & velocity of phase $d \in \{ w, n \}$   \\
			$ p_d $  & pressure of phase $d \in \{ w, n \}$   \\
			$ p_c $  &  capillary pressure function  \\
			$ \mu_d $  &  viscosity of phase $d \in \{ w, n \}$   \\
			$ k_{rd} $  &  relative permeability of phase $d \in \{ w, n \}$  \\
			$ K_0 $  & absolute permeability   \\
			$ g $  & gravity   \\
			$ M_d = \frac{k_{rd}}{\mu_d} $  & mobility  of phase $d \in \{ w, n \}$   \\
			$ M = M_w + M_n $  & total mobility   \\
			$ R_d $  & gas constant of phase $d \in \{ w, n \}$   \\
			$ \rho_d^{ref}, p_d^{ref} $  & reference density and pressure of phase $d \in \{ w, n \}$   \\
			$ u = u_w + u_n $  & total velocity   \\
			$ f_d = \frac{M_d}{M} $  & fractional mobility of phase $d \in \{ w, n \}$   \\
			$ u_D $  & velocity defined by Equation \ref{u_D}    \\
			$ \gamma = \frac{M_w M_n}{M} $  & coefficient function   \\
			$ \bar{S}_d $  & effective saturation of phase $d \in \{ w, n \}$  \\
			$ S_{d,r} $  & residual saturation of phase $d \in \{ w, n \}$   \\
			$ p_c^{VG} $  & Van Genuchten capillary pressure function   \\
			$ p_e^{VG}, m^{VG} $  & parameters of Van Genuchten capillary pressure function    \\
			$ p_c^{BC} $  & Brooks-Corey capillary pressure function   \\
			$ p_e^{BC}, m^{BC} $  & parameters of Brooks-Corey capillary pressure function    \\
			$ k_{rd}^{BC} $  & Brooks-Corey relative permeability of phase $d \in \{ w, n \}$    \\
			$ m^{w,BC}, m^{n,BC} $  & parameters of Brooks-Corey relative permeabilities    \\
			$ C_i $  & cell in the voxel grid   \\
			$ x_{C_i} $  & center of cell $C_i$   \\
			$ f_{ij} $  & shared face of cells $C_i$ and $C_j$   \\
			$ n_{ij} $  & normal of face $f_{ij}$   \\
			$ x_{f_{ij}} $  & center of face $f_{ij}$   \\
			$ \Delta x_{ij} $  &  distance of centers $ x_{C_i} $ and $ x_{C_j} $  \\
			$ n_D \in \{ 2, 3 \} $  &  dimension of the voxel grid  \\
			$ \Delta x_{i,d} $  &  size of cell $C_i$ in dimension $d = 1,...,n_D$   \\
			$ |C_i| $  & area ($n_D = 2$) or volume ($n_D = 3$) of cell $C_i$  \\
			$ e_d $  & canonical unit vector in dimension $d = 1,...,n_D$  \\
			$ \mathbb{V} $  & set of all cells of the voxel grid  \\
			$ F_{C_i} $  & set of all faces of $C_i$  \\
			$ F^d_{C_i} $  & set of all faces of $C_i$ orthogonal to $e_d$   \\
			$ |f_{ij}| $  &  length ($n_D = 2$) or area ($n_D = 3$) of face $f_{ij}$ \\
			$ \bar{\psi}^{H,f_{ij}} $  & harmonic average of $\psi$ at face $f_{ij}$, Eq. \ref{eq:darcySol_harmMeanFace}  \\
			$ \bar{\psi}^{A,f_{ij}} $  & arithmetic average of $\psi$ at face $f_{ij}$, Eq. \ref{eq:darcySol_arithMeanFace}  \\
			$ m_I $  & number of IMPES iterations   \\
			$ \psi_i^k $  &  variable $\psi$ at cell $C_i$ and timestep $k$ \\
			$ \psi_{f_ij}^k $  & variable $\psi$ at face $f_{ij}$ and timestep $k$  \\
			$ F^{up, k_{l-1}} $  &  Eq. \ref{numericalFluxFct} \\
			$ \tilde{k}_l $  &   Eq. \ref{eq:Tildek_l} \\
			$ f_{w, f_{ij}}^{up, k_{l-1}} $  &  Eq. \ref{upwinding_fw}  \\
			$ \tilde{u}^{k_{l-1}}_{D, f_{ij}} $  & Eq. \ref{darcySol_numFluxFct}   \\
			$ u_R $  & Eq. \ref{u_R}    \\
			$ A^{k_l}_{f_{ij}} $  &  Eq. \ref{discretA} \\
			$ B^{k_l}_{f_{ij}} $  &  Eq. \ref{discretB} \\
			$ \hat{u}^{k_l}_{R, f_{ij}} $  &  Eq. \ref{discretu_R} \\
			$ \hat{u}^{k_l}_{D, f_{ij}} $  & Eq. \ref{discretu_D}  \\
			$ C_{\text{stab}} $  &  stability constant  \\
			$ \omega_{ij} $  & characteristic wave velocity at face $f_{ij}$  \\
			$ \omega^{k,max}_{ij} $  & Eq. \ref{eq:darcySol_upperBoundCharWaveVel}  \\
			$ I_{ij}^k $  & Eq. \ref{intervalMax}  \\
			$ \Delta t_k^{gcw} $  & timestep size of the generalized characteristic wave velocity criterion  \\
			$ \tau_{\text{max}} $  & maximal relative increase of timestep size in one time iteration   \\
			$ \Delta_s^{\text{min}} $  & minimal difference of saturations (spatial) in Eq. \ref{velDerivSat}  \\
			$ \Delta_t^{\text{min}} $  & minimal difference of saturations (temporal) in Eq. \ref{velDerivSat}  \\
			$ D^{u,k-1}_{f_{ij}} $  & Approximation of Eq. \ref{velDerivSat}   \\
			$ D^{u_D,k-1}_{f_{ij}} $  & Approximation of Eq. \ref{velDerivSat} but with $u_D$ instead of $u$ 
			
		\end{longtable}

		\section*{Statements \& Declarations}
		\textbf{Author Contributions:}
		All authors contributed to the conception and design of the work. Dominik Burr developed the methodology, implemented the methods, and performed and analyzed the numerical simulations. Konrad Steiner and Stefan Rief supervised the work. Dominik Burr wrote the first draft of the manuscript. All authors reviewed and provided feedback on the first and subsequent versions of the manuscript. All authors read and approved the final manuscript. \\ \\
		\textbf{Conflicts of Interest:}
		The authors have no relevant financial or non-financial interests to disclose. \\ \\
		\textbf{Availability of Data, Code \& Protocols:}
		The developed code of this work cannot be publicly shared as it is part of a commercial software product. \\ \\
		\textbf{Acknowledgements \& Funding:}
		We acknowledges financial support from the High Performance Center Simulation and Software Based Innovation through a PhD scholarship. The authors also acknowledge the financial support provided by Leibniz Collaboration Excellence via the project Machine Learning for Simulation Intelligence in Composite Process Design. No additional funds, grants, or other support were received during the preparation of this manuscript.


\end{document}